%% file: main.tex
\documentclass[acmsmall,authorversion,screen]{acmart}
\acmJournal{PACMHCI}

\AtBeginDocument{%
  }

\usepackage{acmart-taps}
\usepackage{hyphenat}
\usepackage{soul}  
\usepackage[utf8]{inputenc} 
\usepackage{xcolor} 
\usepackage{soul} 
\usepackage{todonotes}
\usepackage{multirow} 
\usepackage{subcaption} 
\usepackage{makecell} 
\usepackage{colortbl} 
\usepackage{fontawesome5} 
\usepackage{csquotes} 
\usepackage{array}
\usepackage[commandnameprefix=always]{changes}
\usepackage{hyperref}
\usepackage{framed} 

\newcommand{\displayquoting}[2]{\begin{displayquote}``\emph{#1}'' (#2)\end{displayquote}}

\definecolor{calloutbg}{rgb}{0.945,0.965,0.984}     
\definecolor{calloutbar}{RGB}{31,120,180}           
\definecolor{indigo}{HTML}{4B0082}     
\definecolor{plum}{HTML}{6A2C7E}        
\definecolor{raspberry}{HTML}{A6366D}  
\definecolor{crimsonred}{HTML}{B22222}  
\definecolor{burntorange}{HTML}{B84A1F} 
\definecolor{goldenrod}{HTML}{8A6914}   
\definecolor{olive}{HTML}{5C5C00}       
\definecolor{forestgreen}{HTML}{1F6F1F} 
\definecolor{emeraldgreen}{HTML}{1E7268} 
\definecolor{teal}{HTML}{007377}       
\definecolor{deepblue}{HTML}{2F4B7C}    
\definecolor{navyblue}{HTML}{003F5C}    
\definecolor{slate}{HTML}{4F5E6C}       
\definecolor{darkgray}{HTML}{555555}    
\definecolor{black}{HTML}{000000}      
\definecolor{fierypurp}{HTML}{5f0f40}   
\definecolor{fieryred}{HTML}{2596be}
\definecolor{fieryblue}{HTML}{d99781}
\definecolor{fieryoj}{HTML}{3e8c77}

\newenvironment{callout}{%
  \MakeFramed{\advance\hsize-\width \FrameRestore}%
  \noindent\ignorespaces
}{\endMakeFramed}

\begin{document}

\title[A Glimpse Into AAA Game Processes and How UX Leaders Navigate Pre-Production]{Theory, Experience, and Instinct: A Glimpse Into AAA Game Processes and How UX Leaders Navigate Pre-Production}

\author{Ivana Randelshofer}
    \authornote{Author is affiliated with the Department of Digital and Analytic Sciences, University of Salzburg}
    \orcid{0009-0009-4597-0418}
    \affiliation{%
    \institution{Ubisoft D\"{u}sseldorf GmbH}
    \city{D\"{u}sseldorf}
    \country{Germany}}
    \email{iva@randelshofer.eu}
    \correspondingauthor

\author{Joseph Tu}
    \authornote{Authors are also affiliated with the Department of Systems Design Engineering, University of Waterloo}
    
    \orcid{0000-0002-7703-6234}
    \affiliation{
    \department{Stratford School of Interaction Design and Business}
    \institution{University of Waterloo}
    \city{Waterloo}
    \country{Canada}}
    \email{joseph.tu@uwaterloo.ca}
    \correspondingauthor 

\author{Yifan Cao}
    \orcid{0000-0002-5892-5052}
    \affiliation{
    \department{University of Science and Technology}
    \institution{Hong Kong University of Science and Technology}
    \city{Hong Kong}
    \country{China}}
    \email{caoyifan@ust.hk}
  
\author{Reza H. Mogavi}
    \orcid{0000-0002-4690-2769}
    \affiliation{
    \department{Department of Computing and Software Engineering}
    \institution{McMaster University}
    \city{Hamilton}
    \country{Canada}}
    \email{mogavir@mcmaster.ca}

\author{Ville M\"{a}kel\"{a}}
    \authornotemark[2]
    \orcid{0000-0001-6095-2570}
    \affiliation{%
    \department{Stratford School of Interaction Design and Business}
    \institution{University of Waterloo}
    \city{Waterloo}
    \country{Canada}}
    \email{ville.makela@uwaterloo.ca}

\author{Lennart E. Nacke}
    \authornotemark[2]
    \orcid{0000-0003-4290-8829}
    \affiliation{
    \department{Stratford School of Interaction Design and Business}
    \institution{University of Waterloo}
    \city{Waterloo}
    \country{Canada}}
    \email{lennart.nacke@acm.org}

\renewcommand{\shortauthors}{Randelshofer et al.}

\begin{abstract}
Foundational decisions shape a project’s long-term trajectory, a dynamic that becomes especially evident in the inherent complexity of game pre-production. However, academic frameworks often see limited uptake at this stage, as they do not readily map onto industry contexts, production constraints, and cross-functional workflows. To better understand how design decisions are made in practice, we conducted interviews with 15 UX leaders from the AAA (triple-A) games industry. Our findings show that early UX decisions emerge from a dynamic blend of theory, experience, and intuition. In cross-functional structures (such as strike and competency teams), UX leaders collaboratively align player needs, technical feasibility, and creative vision. These decision-making processes involve translating academic concepts into production-ready insights, codifying experiential knowledge into reusable practices, and relying on informed intuition amid uncertainty. We argue that meaningful impact requires academia to develop malleable conceptual tools that integrate with practitioners’ highly adaptive design processes. We conclude by discussing how existing frameworks might be adapted to connect academic insights with AAA workflows. Rather than prescriptive directives, we offer these as starting points for discussion that support strike and competency teams through shared language, reusable design systems, and strategies for collaborative, context-sensitive decision-making.
\end{abstract}



\setcopyright{cc}
\setcctype{by}
\acmJournal{PACMHCI}
\acmYear{2026} \acmVolume{10} \acmNumber{7} \acmArticle{GAMES071}
\acmMonth{11} \acmDOI{10.1145/3831330}

\received{February 2026}
\received[revised]{June 2026}
\received[accepted]{July 2026}

\maketitle
\input{sections/00_Introduction} %
\input{sections/01_RelatedWork} %
\input{sections/02_Methods} %
\input{sections/03_Results} %
\input{sections/04_Disc} %
\input{sections/05_Conclusion} %

\begin{acks}
We thank the Games Institute for generously providing the space necessary for this work, and we are grateful the experts from the game industry for their time and involvement in this study. This study was supported by the NSERC Discovery Grant (RGPIN-2023-03705), and the CFI John R. Evans Leaders Fund (CFI JELF-41844). Proofreading and grammar software, including Grammarly (with AI features) and Claude, assisted in restructuring some sentences and reducing word count to improve readability.  Lastly, we sincerely thank our reviewers for their thoughtful suggestions. \textbf{Supplementary material} is available on the Open Science Framework at \url{https://osf.io/mn8e6/overview}.
\end{acks}

\bibliographystyle{ACM-Reference-Format}
\bibliography{references}

\appendix
\input{sections/06_Appendix}
\end{document}

%% file: sections/00_Introduction.tex
\section{Introduction}\label{intro}
Designing and developing video games is a complicated and interdisciplinary endeavor \cite{engstrom2020game,whitson2018}. As video games have grown into a massive industry, they have also gained much attention from researchers over several decades. In particular, researchers have proposed many user experience (UX) theories and practices intended to help game developers (either specifically tailored for games or broader practices that are applicable to games), ranging from design frameworks and principles \cite{Duarte2017DistinctiveFeatures,hunicke2004mda,o2019game} to evaluation methods \cite{bernhaupt2015user,Hochleitner2015,VandenAbeele2020PlayerExperienceInventory}. However, game production is messy and hectic, especially in the early stages (i.e., pre-production) \cite{engstrom2020game,whitson2018}, which raises questions about how game practitioners actually approach game development and decision-making, and whether and how they integrate research theories and practices into early \textit{design stages}. The pre-production phase is critical because it shapes the final product; key decisions about genre, platform, and audience are made then \cite{Chandler2009GameProductionHandbook}, and a lack of UX considerations can lead to costly mistakes that are difficult to fix later \cite{MitreHernandez2016UserExperienceManagement}. Therefore, there is a need to understand current games industry practices during pre-production, and what challenges practitioners face in integrating academic work into their production phase. 

In this paper, we use ``UX'' as an umbrella term encompassing user experience design, user interface design, games user research (GUR), and related roles concerned with how players engage with games. We focus on \textit{UX leaders}: professionals in these roles who define the user experience vision, align design with business goals, and coordinate cross-disciplinary teams to create player-centric products. This understanding is critical for both academia and industry. First, research contributions should strive to make real-world impact \cite{Bornmann2013SocietalImpact,Graham2018MovingKnowledge,Perkmann2013AcademicEngagement}, and understanding the pitfalls, bottlenecks, and realities of the relevant industries will help academia generate and revise applicable theories and practices \cite{o2019game,karlsson2023level}. Second, the games industry faces significant challenges in production and management \cite{Petrillo2008Houston,Petrillo2009,politowski2021game}, including chaotic production pipelines \cite{whitson2018} and limited integration of UX considerations during development \cite{MitreHernandez2016UserExperienceManagement}. These structural challenges shape how---and whether---academic insights translate into design practice.

To address this, we examine how UX industry professionals navigate design decisions in the early stages of game production, including ideation, cross-disciplinary collaboration, current design strategies, and the barriers practitioners face in adopting formal research methods. We conducted semi-structured interviews with 15 lead, senior, and director-level UX and UI professionals from AAA (triple-A)\footnote{We adopt the conventional definition of AAA (triple-A) as the segment of the video game market that produces high-budget (with budgets upward of 50 million US dollars), high-production-value games, characterized by extensive development teams (100+ employees per team) and cutting-edge technology~\cite{alagappa2021aaa}.} game companies. Game studios tend to protect their intellectual property (IP) against competitors, which makes it hard for researchers to access these experts \cite{ruggill2016inside}, thereby making our insights rare and valuable. We intentionally focused on AAA productions, as their scale, complexity, and resource-intensive nature provide a rich context for understanding both the opportunities and constraints practitioners encounter. Further, we focused on senior positions---including directors and leads---because they have oversight over large teams and processes as well as strategic influence over UX decisions in high-budget, high-stakes environments \cite{Martin2018Intellectual,svelch2021}. Specifically, our study addresses three research questions:

\begin{itemize}
    \item [] \textbf{RQ1:} How do AAA UX design industry practitioners approach ideation and creation during early game development?
    \item [] \textbf{RQ2:} What structural, methodological, or theoretical gaps prevent AAA UX design industry practitioners from effectively incorporating academic practices?
    \item [] \textbf{RQ3:} What strategies could enhance the applicability of academic practices for industry use in game development?
\end{itemize}

From this, we identify three guiding principles for decision-making in pre-production: aca\-dem\-i\-cally-grounded, experience-based, and gut-feeling-driven approaches. Our findings indicate that early UX decision-making in game development is shaped by a blend of theory, experience, and intuition. In cross-functional roles, strike and competency teams work together to align design with player needs, technical constraints, and the overarching creative vision. Rather than rejecting academia, experts highlighted the need for adaptable, context-sensitive approaches that align with critiques of rigid ``one-size-fits-all'' methodologies in HCI. However, tight schedules, stakeholder demands, and limited awareness of applicable research frameworks often restrict research use; research is most applicable in pre-production, yet frequently serves as justification rather than guidance. In summary, our primary contributions are: 

\begin{enumerate}

    \item \textbf{Empirical Contribution (RQ1, RQ2)}: We surface tensions between academic approaches and industry realities as described by senior AAA UX leaders, outlining the barriers they identify to research adoption and the practical methods they use to navigate decision-making. From this, we identify three guiding principles practitioners describe---academically-grounded, experience-based, and gut-feeling-driven approaches---which we synthesize into a descriptive Model of Adaptive Design Judgment (ADJ).
    \item \textbf{Practical Implications (RQ3)}: We discuss how academic practices might be aligned with industry timelines---especially pre-production---and real-world production constraints. Participants described how their teams already capture and standardize design knowledge in shared artifacts such as design systems and internal guidelines. Because these artifacts structure everyday design work, they may offer practical points at which research and practice can be connected. We offer these as starting points for discussion rather than prescriptive directives, including possibilities for providing concrete use cases, reusable components, and clearer guidance that complements existing industry practices.
    
\end{enumerate}

To understand how and why this disconnect persists, we next review previous research on game production processes, interdisciplinary challenges, and the theory-practice gap in games UX.

%% file: sections/01_RelatedWork.tex
\section{Related Work}\label{related_work}
In this section, we review previous work on game production, processes and pipelines related to game development. Despite this body of work, a persistent disconnect remains between academic accounts of interdisciplinary workflows and production challenges, and the ways these are experienced and managed in practice.

\subsection{Processes and Pipelines in Game Development}
Game development pipelines are dynamic and interdisciplinary~\cite{politowski2021game,guevara2011cultures}, with fluid roles~\cite{karlsson2020investigating} that demand reactive processes~\cite{ernkvist2018differentiation,jorgensen2019newcomers}, making standardized workflows challenging~\cite{kanode2009software,ramadan2013,aleem2016game}. Such pipelines require coordination across diverse disciplines—art, engineering, and design \cite{Musil2010}. This complexity creates challenges in aligning workflows, from requirement gathering to multi-platform development. \citet{Birdwell2006} emphasizes that successful game development depends on cross-disciplinary collaboration to achieve high-value outcomes.

\definecolor{starGold}{HTML}{DAA520}
\definecolor{focusHighlight}{HTML}{f0f0f0}  

\subsubsection{Game Development Life Cycle (GDLC)}

The Game Development Life Cycle consists of six stages tailored to gaming's unique demands, including artistic iteration and gameplay-centred evaluation~\cite{ramadan2013}. \autoref{tab:game-dev-stages} outlines these stages based on \citet{Atkas2014} and \citet{crawford2003chris}.

\begin{table}[ht]
    \centering
    \footnotesize
       \begin{tabular*}{\columnwidth}{@{\extracolsep{\fill}}p{0.35\columnwidth}p{0.6\columnwidth}@{}}
        \toprule
        \textbf{Stages} & \textbf{Description} \\
        \midrule
        Specification (or Initiation) &
        Defining key features of the game, including target audience, genre, high concept documents, platform specifications, and a high-level development plan. Considered the most creative part of the development process. \\
        \midrule
        {\color{starGold}\faStar}~Pre-production (or Early-Development) &
        Development of game and level design concepts, story, essential game mechanics, aesthetics, development principles, object-oriented specifications, and the design of the game world. \\
        \midrule
        Production (or Development) &
        Focused on crafting and delivering content, such as computer code, models, sounds, videos, and their integration into the game. \\
        \midrule
        Post-Production (or Validation and Testing) &
        Driven by Quality Assurance (QA), focusing on discovering issues in code, game mechanics, gameplay, user interfaces, and audio-visual content to meet market requirements. \\
        \midrule
        Release and Launch &
        Delivery of the release to the manufacturer. \\
        \midrule
        Post-release (or Maintenance) &
        Ongoing improvement of the product through patches and upgrades. \\
        \bottomrule
        \end{tabular*}
        \caption{\small Typical Stages of Game Development Life Cycle (GDLC) based on \citet{Atkas2014} and \citet{crawford2003chris}. Our study focuses on the \textit{Pre-production} stage (denoted with {\color{starGold}\faStar} in the table).}
        \Description{Table 1 presents the typical game development life cycle, which is divided into six stages in total. Stage one is Specification, stage two is Pre-production, stage three is Production, stage four is Post-production, stage five is Release and Launch, and stage six is Post-release. The table compares the summary of these stages as presented by Aktas and Orcun with a newly edited version by Crawford and colleagues.}
        \label{tab:game-dev-stages}
\end{table}

Game development comprises core areas, including graphics, sound, game logic, physics, scripting (development), design, and user interfaces \cite{Sobota2023}. Expertise in user interface (UI) and user experience (UX) design is essential for intuitive player-game interaction, while interpersonal skills and team collaboration are crucial for multidisciplinary teams~\cite{Sturdee2022}. \citet{Fidas2015} note that development requires different stakeholders with distinct roles that must be balanced throughout production. However, most work focuses primarily on player experiences rather than developer perspectives and practices~\cite{lucy2021,denoo2025developer}, and rarely engages with typical GDLC realities. This gap is most pronounced in \textbf{pre-production} (see~\autoref{tab:game-dev-stages}), which centres on concept validation, prototyping, and risk reduction by small teams, in contrast to the full-scale content creation, large teams, and significant resource investment of the \textbf{production} stage~\cite{crawford2003chris,Atkas2014}. Our study focuses specifically on pre-production, where UX strategy is shaped and embedded before large-scale resources are committed.

\subsection{Gaps Between Academia and Games Industry}
Academia has long been interested in understanding the game industry's processes. As such, empirical research has previously investigated the disconnect between research theory and industry practices~\cite{guevara2011cultures,Sturdee2022,baharom2014emotional}. \citet{whitson2018} draws attention to the gap between scholars' conceptualizations of game production and the ``messy realities'' of collaborative development. They outline substantial discrepancies between research expectations and the actual challenges developers face, rooted in constant cross-disciplinary negotiation. Foundational design textbooks~\cite{Bjork2005,Fullerton2008,Koster2005} and development studies \cite{Petrillo2009,Shirinian2011,Ash2015} typically emphasize technical processes and formal methodologies, while practitioner accounts \cite{Banks2013,Malaby2009} focus on studio workflows. However, \citet{ODonnell2014} shows how informal ``game talk'' enables interdisciplinary collaboration by helping developers navigate both technical systems and social dynamics.

Building on this perspective, \citet{Velt2020} highlights the persistent gap between HCI research and UX practice, attributing it to tensions between high-level theoretical contributions and the realities of low-level design work~\cite{Gaver2012,Rogers2004,Stolterman2008}, as well as to differing community priorities, such as generating generalizable knowledge versus crafting situated, context-specific solutions~\cite{Nelson2012,Goodman2011}. This disconnect is further compounded by limited transparency in professional practice; for instance, \citet{karabinus2018games} notes that ``professional game design processes and practices are often obfuscated, [making] it difficult for researchers to study how game design happens.'' 

At the same time, much of the existing research on game structure draws on insights from UX participants who do not hold decision-making authority to substantially shape that structure, a dynamic also reflected in the International Game Developers Association (IGDA)’s Developer Satisfaction Survey (DSS)\footnote{IGDA Developer Satisfaction Survey Reports: https://igda.org/dss/}. This challenge is further compounded by the commercial game industry’s volatility, as evidenced by recurring waves of layoffs that have destabilized development environments, as noted in the DSS Summary Report of 2023. Together, these factors highlight the need to translate HCI knowledge more effectively into actionable forms (particularly intermediate- or higher-level design knowledge), such as heuristics and guidelines that can better inform practical design decisions~\cite{Lowgren2013}. However, little is known about how such knowledge is interpreted, adapted, and operationalized within large-scale commercial contexts, particularly by UX leaders\footnote{We use UX leaders as an umbrella term for senior gaming-industry professionals---including UX/UI designers, games user researchers, and related specialist roles---who define the user experience vision, align design with business goals, and coordinate cross-disciplinary teams to create player-centric products. Although this term may carry different meanings across companies, we use it consistently as defined here throughout the paper.} in AAA game studios who must mediate between research insights, production constraints, and cross-disciplinary teams.

\subsection{Knowledge Translation, Tacit Practice, and Design Thinking}
Several theoretical traditions relevant to our study have examined how practitioners bridge research and practice. The most directly applicable to HCI is \citet{colusso2019translational}'s mapping of the \textit{Applied Research to Design Practice} (TAD) gap, which proposes a translational continuum from basic science to design tools. This framing aligns with broader characterizations of the same problem---\citet{Norman2010}'s account of the research-practice gap as structural, in which findings take years to reach practitioners and rarely arrive in actionable form, and \citet{Graham2018MovingKnowledge}'s \textit{know-do gap}, where valuable knowledge fails to change practice. \textit{Integrated knowledge translation} (iKT) models propose sustained researcher-practitioner co-production as a remedy~\cite{Graham2018MovingKnowledge}, but such depth of collaboration is rarely feasible in commercial game production.

\citet{star1989institutional} introduced \textit{boundary objects} as artifacts that are flexible enough to be interpreted across communities yet robust enough to maintain a shared identity. In interdisciplinary game development, design systems, heuristics, and shared vocabularies can serve as objects that enable coordination among UX, engineering, and creative teams without requiring consensus on an underlying theory. In addition to this, practitioners rely heavily on \textit{tacit knowledge} (i.e., expertise that is difficult to articulate or transfer formally~\cite{wong2000tacit,schindler2015expertise}). \citet{schon_reflective_2017}'s concept of the \textit{reflective practitioner} describes how expert designers navigate uncertainty through \textit{reflection-in-action}, where they continuously reframe problems as they work. These perspectives suggest that practitioner expertise cannot be fully captured by formal frameworks alone and that codifying experiential knowledge into reusable artifacts is itself a significant design activity.

Design Thinking (DT) has been proposed as a bridge between structured research methods and the exploratory and iterative realities of design practice~\cite{seidel2013adopting, lockwood2009transition}. DT treats problems as open-ended and supports teams in navigating the fuzzy front end of product development through iterative cycles of needs finding, ideation, and prototyping~\cite{cooper2009design,de2021acquaintances}. Although DT is rarely formalized in AAA contexts, its principles (particularly tolerance for ambiguity, iterative refinement, and collaborative sense-making) resonate with practitioners' descriptions of pre-production work. However, DT's applicability in large-scale, milestone-driven environments remains underexplored.
Across these traditions, our stance is interpretive rather than evaluative. We draw on knowledge-translation models, accounts of tacit and reflective practice, and Design Thinking not as frameworks to test or validate, but as analytic lenses for understanding how UX leaders in AAA studios reconcile research insight with the production realities of large-scale commercial development.

Despite this body of work, three specific gaps remain unaddressed. \textbf{First} (RQ1), it is unknown how UX leaders in AAA studios actually blend academic theory, codified experience, and informed intuition in early design decision-making, and whether these approaches align with reflective practice or Design Thinking models. \textbf{Second} (RQ2), the structural barriers that prevent knowledge translation in large-scale game production have not been empirically examined at the leadership level. \textbf{Third} (RQ3), practical strategies for making academic frameworks translation-ready within AAA workflows, including how boundary objects such as design systems can mediate between research and practice, remain largely unspecified. Our study addresses all three gaps through semi-structured interviews with senior UX leaders in AAA game studios.

%% file: sections/02_Methods.tex
\section{Methodology}\label{methods}
This study was approved by the University of Waterloo's Research Ethics Board (REB \#45589), ensuring adherence to ethical research guidelines, including informed consent and participant confidentiality. After the study, participants received a thank-you letter as a token of appreciation for their time and contributions; no monetary compensation or other remuneration was provided.

\subsection{Recruitment}
We recruited 15 UX leaders from AAA game companies by posting recruitment calls on social media platforms and through snowball sampling, in which participants referred eligible colleagues. Interested respondents completed a screening questionnaire that collected their current role, relevant experience and interest, company size, and the primary purpose of their game design work. We screened responses against our eligibility criteria: participants had to be at least 18 years old, have a stable internet connection and access to a computer or other internet-connected device for the online interview, have experience crafting/designing, researching, or evaluating game user interfaces (UIs) in the games industry, and consent to having the interview audio-recorded. In practice, the experience criterion served as a screener for seniority, as the authority to evaluate and shape game UIs---rather than merely execute predefined designs---is typically entrusted to lead-, senior-, and director-level practitioners. We then contacted eligible respondents via email to invite them to participate in semi-structured online interviews.

\subsection{Expert Participants}
All participants held lead-, senior-, or director-level positions in game UX design or UI design, and currently work at, or have previously worked at, major game companies including Blizzard Entertainment, Electronic Arts Digital Illusions CE (EA DICE), Guerrilla Games, Larian Studios, Remedy Entertainment, and Ubisoft, spanning titles across multiple genres and platforms. We deliberately recruited at this level of seniority because senior game UX practitioners constitute a small professional community, and detailed individual profiles would pose a re-identification risk; in line with our ethics approval and participant consent, we report roles and experience in generalized form (\autoref{tab:participant}) and do not link any individual participant to a named company.

Participants' responsibilities included defining user experience vision, aligning design with business goals, and empowering teams to create user-centric game products. This seniority afforded them strategic influence over design processes---coordinating cross-disciplinary collaboration, identifying gaps, and devising mitigation strategies~\cite{alves2007challenges}---making them uniquely positioned to integrate academic practices into daily workflows. For similar reasons, we concentrated on practitioners from AAA studios. While smaller studios and indie companies often lack the resources to sustain extended processes or employ specialized experts, AAA studios typically maintain large, specialized teams and substantial financial resources, supporting investment in sophisticated workflows and in-house tools~\cite{bernevega2022industry, berg2019empirically}. This focus allows us to examine how academic practices are integrated into large, complex team structures and pre-production pipelines.

\begin{table}[!b]
    \centering
    \small
    \begin{tabular*}{\columnwidth}{@{\extracolsep{\fill}}m{0.2\columnwidth}m{0.25\columnwidth}m{0.25\columnwidth}m{0.25\columnwidth}@{}}
        \toprule
        \textbf{Expert ID} & \textbf{Role} & \textbf{Experience$^a$}  & \textbf{Duration$^b$}               \\
        \midrule
        \color{indigo}\faUser~Expert 1  & UI Lead   &  20 years+  & 57\,min                     \\
        \color{plum}\faUser~Expert 2  & UX Content Director  & 20 years+  & 75\,min             \\
        \color{raspberry}\faUser~Expert 3  & UI/UX Lead  & 10 years+  & 43\,min                 \\
        \color{crimsonred}\faUser~Expert 4  & Senior UX Designer$^*$ &  5 years+ & 70\,min           \\
        \color{burntorange}\faUser~Expert 5  & Creative/Game Director  &  15 years+ & 70\,min        \\
        \color{goldenrod}\faUser~Expert 6  & UI/UX Lead  &  25 years+ & 51\,min                 \\
        \color{olive}\faUser~Expert 7  & Senior Product Designer$^*$ & 5 years+  & 40\,min         \\
        \color{forestgreen}\faUser~Expert 8  & UI/UX Director &   25 years+  & 63\,min         \\
        \color{emeraldgreen}\faUser~Expert 9  & Design Lead  & 5 years+    & 52\,min           \\
        \color{teal}\faUser~Expert 10 & Senior UX Designer$^*$ &  15 years+  & 45\,min             \\
        \color{deepblue}\faUser~Expert 11 & UX Director   &  10 years+ & 41\,min              \\
        \color{navyblue}\faUser~Expert 12 & UX Director  &  25 years+   & 58\,min              \\
        \color{slate}\faUser~Expert 13 & UX Director   &  25 years+   & 81\,min                 \\
        \color{darkgray}\faUser~Expert 14 & UX Director     &  15 years+ &  56\,min     \\
        \color{black}\faUser~Expert 15 & Senior UI Artist$^*$ &  5 years+  & 71\,min                  \\
        \bottomrule
    \end{tabular*}
    \caption{Overview of industry experts and their respective game development roles.  \\ $^a$ Years of experience are self-reported and grouped into ranges to preserve anonymity. \\ $^b$ Duration of the semi-structured interviews rounded to the nearest full minute (min). \\ $^*$ As senior practitioners, these participants are subject-matter experts (SMEs) who \textbf{lead} the development of important features, including impactful decisions during pre-production; thus, despite not having people-management duties, they are considered leaders in their area of expertise.}
    \Description{Table 2 provides an overview of 15 industry experts, describing their roles and levels of experience within their respective companies. The years of experience are self-reported and presented in ranges, from five years to more than twenty-five years, in order to preserve anonymity. All experts included in this table work within the AAA game industry, with roles that vary from UI and UX Leads to Directors of UX.}
    \label{tab:participant}
\end{table}

\subsection{Positionality Statement}\label{positionality}

Our analysis was shaped by the team's individual and collective backgrounds. Following reflexive thematic analysis, we treat these as resources for interpretation rather than as biases to be eliminated~\cite{may2014reflexivity, Braun2021Size, braun2023doing}. The first author brings over 25 years of industry experience in UX, research, and game development of AAA games; this insider perspective informed how we framed initial codes around production realities and shaped tentative theme development. The second author also has games industry experience designing hybrid game systems and has played all of the games discussed by participants, which directly shaped our interpretation of participant references and industry slang that the rest of the team might otherwise have missed. The third author's expertise in UX visualization sensitized the analysis to themes concerning practitioners' visual artifacts and tooling. The remaining team members brought HCI perspectives that grounded the analysis in broader theoretical frames and mediated between the industry-insider interpretations of the first three authors and academic framing. The full team collaborated on rephrasing themes for clarity.

The research team represents a range of professional and academic perspectives, from experienced UX researchers in the games industry to graduate students and full professors. Our collective backgrounds span games user research (GUR), computer science, cognitive systems, interaction design, and user experience (UX). As part of our reflexive approach, we also considered how our individual experiences with AAA games (including role-playing games, first-person shooters, survival horror, couch co-op, action, and turn-based strategy) shaped our interpretations and sensitivity to the data from an industry and researcher perspective. Throughout coding and theme development, we actively reflected on how these positions shaped what we noticed, foregrounded, and interpreted in the data.

\begin{figure}
    \centering
    \includegraphics[width=0.6\linewidth]{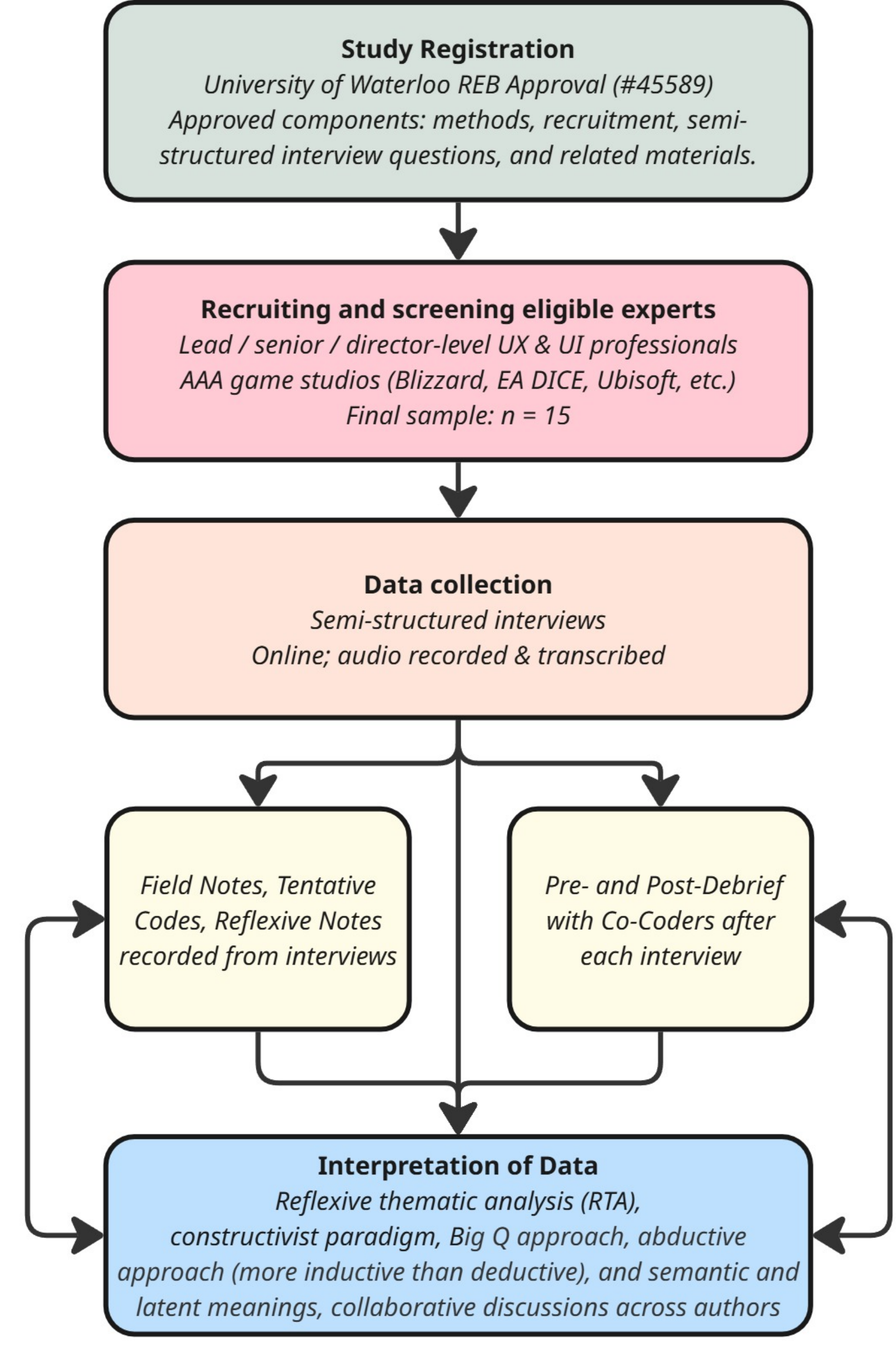}
    \caption{Study methodology flowchart illustrating the research process: study registration, expert recruitment and screening, data collection via semi-structured interviews, with parallel field noting and co-coder debriefing feeding into iterative interpretation of data using reflexive thematic analysis.} 
    \Description{A vertical flowchart illustrating the study methodology in six boxes. Box one, Study Registration, describes University of Waterloo REB approval and lists approved components including methods, recruitment, semi-structured interview questions, and related materials. Box two, Recruiting and Screening Eligible Experts, describes the participant criteria: lead, senior, or director-level UX and UI professionals from AAA game studios including Blizzard, EA DICE, and Ubisoft, with a final sample of 15 participants. Box three, Data Collection, describes semi-structured interviews conducted online, audio recorded and transcribed. From Data Collection, two parallel branches emerge: the left branch captures field notes, tentative codes, and reflexive notes recorded from interviews; the right branch describes pre- and post-debriefs with co-coders after each interview. Both branches feed into the final box, Interpretation of Data, which describes the analytical approach as reflexive thematic analysis using a constructivist paradigm, Big Q approach, abductive approach (more inductive than deductive), and attention to semantic and latent meanings through collaborative discussions across authors. Feedback arrows loop between the two branch boxes and Interpretation of Data, indicating an iterative process.}
    \label{fig:procedure}
\end{figure}

\subsection{Data Collection}
The first author conducted all semi-structured interviews. The audio recordings were transcribed using \textit{Dovetail}\footnote{Dovetail is an online platform for analyzing qualitative data, allowing users to tag and code transcriptions to identify patterns and insights.}. Throughout data collection, the first author maintained reflexive notes (or memos) capturing observations on participants' collaboration and workflow, design processes, design principles and frameworks, and continuous learning practices, as well as industry-related context that may have shaped each conversation. Because semi-structured interviews gave the first author latitude in probing, phrasing, and pursuing emergent topics, pre- and post-interview debriefs with the second and third authors prompted reflection on how these in-the-moment choices shaped the resulting data. For example, after the third interview, the focus shifted as richer dynamics emerged around design systems and team structures, and subsequent interviews probed these more deliberately. During transcription and in field notes (on-the-scene recordings), the first author also recorded silences, vulnerable moments, realizations, and other non-verbal cues, attending to how things were said as well as what was explicitly stated; for example, participants often became guarded or hesitant when discussing stakeholder pressure amid ongoing industry layoffs, or when touching on failed games that never reached production and are rarely known to the public, which prompted us to probe these topics with caution.

\begin{figure}[ht!]
    \centering
    \includegraphics[width=0.62\linewidth]{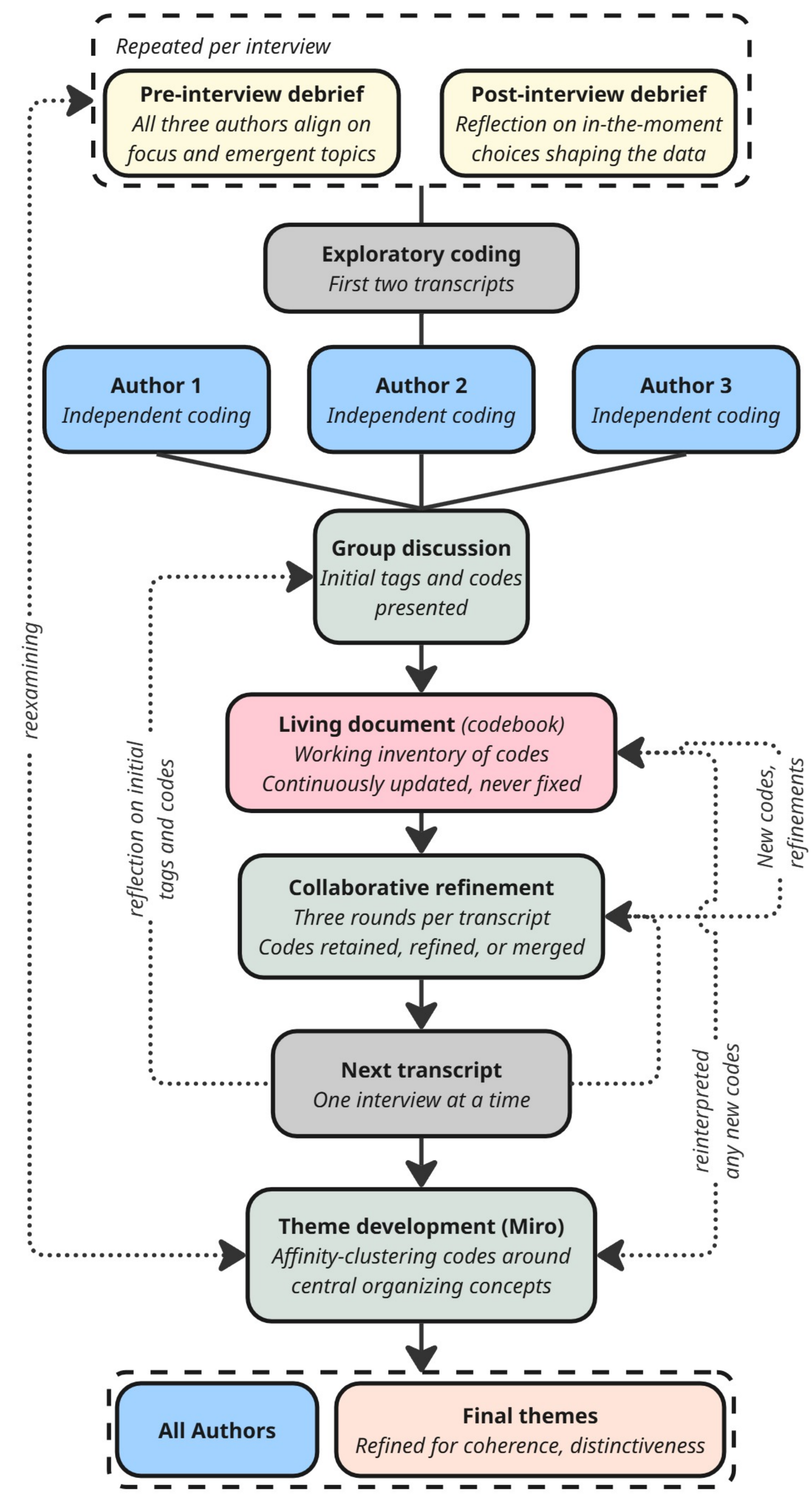}
    \caption{Coding and theme development workflow. Pre- and post-interview debriefs repeated for each interview. All three authors independently coded the first two transcripts before establishing a shared living document (codebook), refining codes across three rounds of dialogue per transcript. Dotted feedback loops reflect our abductive approach. All authors then constructed themes in Miro through affinity-clustering and iterative refinement.}
    \Description{A vertical flowchart of the coding and theme development workflow. At the top, a dashed container labeled Repeated per interview holds two boxes: Pre-interview debrief, where all three authors align on focus and emergent topics, and Post-interview debrief, involving reflection on how in-the-moment choices shaped the data. These feed into Exploratory coding of the first two transcripts, which leads to three parallel boxes for Author 1, Author 2, and Author 3, each conducting independent coding. The three converge into Group discussion, where initial tags and codes are presented, which feeds into the Living document (codebook), described as a working inventory of codes, continuously updated and never fixed. This leads to Collaborative refinement, with three rounds per transcript in which codes are retained, refined, or merged, followed by Next transcript, one interview at a time. Four dotted feedback arrows indicate the iterative, abductive character of the analysis: one loops from Collaborative refinement back to the Living document, labeled new codes and refinements; one loops from Next transcript back to Group discussion, labeled reflection on initial tags and codes; one runs down the right side from the Living document to Theme development, labeled reinterpreted any new codes; and one bidirectional arrow runs along the far left between the debrief stage and Theme development, labeled reexamining, indicating that debrief insights informed theme development and theme development prompted returns to earlier reflections. The flow concludes with Theme development in Miro, using affinity-clustering of codes around central organizing concepts, leading to a final dashed container holding two boxes: All Authors and Final themes, refined for coherence and distinctiveness.}
    \label{fig:coding}
\end{figure}

\subsection{Interpretation of Data}
We analyzed the interview data using reflexive thematic analysis~\cite{braun2023doing}, adopting a constructivist, Big Q orientation to qualitative inquiry: we saw ourselves as active interpreters of the data and treated knowledge as co-constructed between researchers and participants rather than discovered from the data. When we read the data differently from one another, we used those differences to deepen the analysis rather than treating them as mistakes~\cite{braun2023doing}. Our analytical approach was ``abductive'' (more inductive than deductive), combining data-driven coding with sensitivity to existing theoretical concepts, including the game development life cycle~\cite{Atkas2014,crawford2003chris} and Design Thinking~\cite{seidel2013adopting,cooper2009design}. This follows from our constructivist stance: researchers never read data from a blank slate, so our analysis necessarily moved between what participants said and the concepts we brought to it.

The first author cleaned the automated transcriptions and immersed themselves in the data through repeated readings before coding the first two transcripts exploratorily; the second and third authors then independently coded these same transcripts. All authors subsequently met to discuss their initial tags and codes, recorded in a shared ``living document'' (codebook)\footnote{We use the term ``codebook'' pragmatically rather than in its codebook-TA sense. Reflexive thematic analysis is typically distinguished from codebook and coding reliability approaches to TA~\cite{braun2023doing}, so we clarify our usage: in our analysis (shown in~\autoref{fig:coding}), the codebook functioned as a ``living document''---a working inventory of codes used to track, cluster, and refine analytic observations as our understanding developed---rather than as a fixed instrument applied to the dataset or used to check coding consistency. Treating the codebook as a living document is consistent with reflexive TA, where the goal, as \citet{braun2023doing} put it, is to ``engage in analytic process in the way that works best for you and your project, yet remains conceptually coherent''. Accordingly, new codes continued to be generated throughout the analysis, and existing codes were refined or merged as we developed our interpretation of the patterns of meaning in the data.}. Coding proceeded one interview at a time, beginning with simple semantic codes focused on the explicit meaning of what participants had said, and expanding to address latent meanings as the analysis matured. When working through each transcript, the first author consulted the corresponding reflexive and field notes, revisiting and annotating previously coded transcripts in light of new insights. The first and second authors approached coding as ``consciously curious'' researchers: open to hearing (and reporting) experiences different from their own, and seeking both to connect the data to what was already familiar from their industry experience and to make sense of unfamiliar practices described at other companies~\cite{trainor2021developing}. The third author approached coding ``holistically'': attending to patterns across the full dataset rather than within individual transcripts, and questioning interpretations that fit one interview but not the broader patterns. This process (see \autoref{fig:coding}) helped scale down the initial codes and identify those most relevant to the research questions for collective discussion.

The first and second authors shared contextual familiarity with the community-specific language and slang used by participants, which supported reflexive engagement with the latent meanings embedded in informal or insider terminology. For each interview transcript, the three authors discussed the codes across three rounds of collaborative refinement, retaining, refining, or merging codes through dialogue. These discussions often generated new codes and alternative readings of the data rather than converging on a single correct interpretation. Theme development was conducted manually to preserve the interpretive character of the analysis, using Miro\footnote{Miro is a collaborative whiteboard platform that supports distributed teamwork through visual mapping and organization.} to organize and explore patterns across codes. Themes were constructed by affinity-clustering codes around shared central organizing concepts, then iteratively refined for coherence and distinctiveness, with all authors collaborating on final theme development.

\subsubsection{Reflexive Interpretation and Tagging (As an Example)}

\displayquoting{I think you just have to convince the key people, talk to them. I think there's no magic solution for this. [...] I happen to be really good friends with the game director. A lot of late-night beers, basically, and arguing in bars.}{{\color{raspberry}\faUser~Expert 3}}

We interpreted this quote reflexively by attending to both its explicit content and the latent dynamics it described. {\color{raspberry}\faUser~Expert 3}'s account was unpacked reflexively, highlighting their reliance on informal discussions, negotiations, and personal relationships to guide product direction. Although the team previously attempted to introduce a more structured process, it was not broadly adopted. We read this as describing a decision-making culture in which formal process had been attempted and abandoned, leaving informal, relationship-mediated mechanisms---bars, late-night beers, friendship with the game director---as the actual coordination infrastructure. Later in our analysis, we observed that {\color{goldenrod}\faUser~Expert 6} and {\color{teal}\faUser~Expert 10} described similar dynamics as ``gut instinct,'' while {\color{olive}\faUser~Expert 7} and {\color{navyblue}\faUser~Expert 12} referred to it as an internal ``feeling.'' Despite the differences in surface language, we read these as describing a shared underlying phenomenon: informal, embodied, relationship-mediated decision-making in the absence of formal process. Accordingly, we consolidated these extracts under the tag ``gut feeling.'' We note that this represents a coding tag rather than a developed theme (see \autoref{tab:themes} for themes and sub-themes).

%% file: sections/03_Results.tex
\section{Results}\label{results}

\begin{table*}[b!]
    \centering
    \small
\begin{tabular}{p{6cm}|p{7cm}}
\toprule
\textbf{\faListUl~Main Themes} & \textbf{\faChevronCircleRight~Sub Themes} \\ 
\midrule

\rowcolor{fierypurp!10}
\makecell[{{p{6cm}}}]{
\textcolor{fierypurp}{\faCircle}~\textbf{Theme 1:} UX Decision-Making in Early Game Development Is Guided by a Dynamic Mix of Evidence, Experience, and Intuition  
}
&  \makecell[{{p{7cm}}}]{
\textcolor{fierypurp}{\faChevronCircleRight}~\emph{Academically-Grounded, Theory- and Research-Driven Approach} \\ 
\textcolor{fierypurp}{\faChevronCircleRight}~\emph{Experience-Based Approach (Systematic with an Industry Lens)} \\ 
\textcolor{fierypurp}{\faChevronCircleRight}~\emph{Gut Feeling-Driven Approach (Non-Systematic Individual Preferences with Personal Lens)}
} 
\\
\midrule

\rowcolor{fieryred!10}
\makecell[{{p{6cm}}}]{
\textcolor{fieryred}{\faCircle}~\textbf{Theme 2:} Designers Operate in Transversal, Collaborative Roles Across the Production Pipeline to Align Design with Player Needs, Technical Constraints, and Creative Vision 
}
&  \makecell[{{p{7cm}}}]{
\textcolor{fieryred}{\faChevronCircleRight}~\emph{Strike Teams (Cross-Functional, Feature-Focused Units)} \\ 
\textcolor{fieryred}{\faChevronCircleRight}~\emph{Competency Teams (Expertise-Driven, Discipline-Based Clusters)}
}
\\
\midrule

\rowcolor{fieryoj!10} 
\makecell[{{p{6cm}}}]{
\textcolor{fieryoj}{\faCircle}~\textbf{Theme 3:} Stakeholder Priorities and Awareness Gaps Are Key Barriers to Adopting Frameworks in Complex Game Projects 
}
& \makecell[{{p{7cm}}}]{
\textcolor{fieryoj}{\faChevronCircleRight}~\emph{Pre-Production as the Critical Window and the Impact of Production Demands on Design Decisions} \\ 
\textcolor{fieryoj}{\faChevronCircleRight}~\emph{Research Lacks Visibility and Actionable Formats for Industry Application} \\ 
\textcolor{fieryoj}{\faChevronCircleRight}~\emph{Experts Call for Concrete, Contextualized Use Cases}
} 
\\
\midrule

\rowcolor{fieryblue!10}
\makecell[{{p{6cm}}}]{
\textcolor{fieryblue}{\faCircle}~\textbf{Theme 4:} Shared Language and Common Ground Can Enable Stronger Collaboration Between Academia and Practice
}
& \makecell[{{p{7cm}}}]{
\textcolor{fieryblue}{\faChevronCircleRight}~\emph{Design Systems as the Foundation for Shared Vocabularies and Understanding} \\ 
\textcolor{fieryblue}
{\faChevronCircleRight}~\emph{"Conceptual Lego Pieces" as a Basis for Assembling Custom, Modular Frameworks} 
} 
\\
\bottomrule
\end{tabular}

\caption{An overview synthesis of the four themes and subthemes.}
\Description{Table 3 synthesizes four main themes and their associated sub-themes. Theme 1 (pale-red) highlights that UX decision-making in early game development is guided by a dynamic mix of evidence, experience, and intuition, encompassing academically grounded, theory- and research-driven approaches, experience-based systematic methods with an industry lens, and gut feeling-driven approaches that reflect personal, non-systematic preferences. Theme 2 (pale-orange) emphasizes that designers operate in transversal, collaborative roles across the production pipeline to balance player needs, technical constraints, and creative vision, through strike teams—cross-functional, feature-focused units—and competency teams—expertise-driven, discipline-based clusters. Theme 3 (pale-blue) identifies stakeholder priorities and awareness gaps as key barriers to adopting frameworks in complex projects, with critical pre-production windows affected by production demands, limited visibility, and a lack of actionable research formats, and by expert calls for concrete, contextualized use cases. Finally, Theme 4 (pale-green) suggests that shared language and common ground can enable stronger collaboration between academia and practice, supported by modular ``Lego-like'' building blocks and design systems that serve as foundations for shared vocabularies and understanding.}
\label{tab:themes}
\end{table*}

The following section presents findings from semi-structured interviews with UX leaders. These experts frequently referred to UX and UI practitioners as ``Designers,'' a terminology we adopt here to remain consistent with their language. Their design processes are often treated as proprietary endeavors \cite{mcdaniel2016developer}, guarded against other studios that might seek to replicate their methods.

\subsection{\texorpdfstring{\textcolor{fierypurp}{\faCircle}}~Theme 1: UX Decision-Making in Early Game Development Is Guided by a Dynamic Mix of Evidence, Experience, and Intuition}\label{theme_1} 
Early UX decision-making in AAA studios was not governed by any single method. Practitioners instead drew on three distinct logics---formal academic evidence, accumulated professional experience, and intuition---moving fluidly among them according to project context. What distinguished expert practice was not commitment to any one approach but the capacity to blend all three. Experts noted that every project is unique, making it crucial to select and adapt methods to fit each context. They emphasized that the goal is not the wholesale adoption of frameworks, but rather their translation into pragmatic, lightweight tools that respect the realities of production while retaining the rigor of research. 

The challenge, then, is balancing academic rigor with the practical constraints of large‑scale game development. Once assets are integrated into the game engine, iteration becomes costly because changes demand substantial technical work and cross‑team coordination. This aligns with \citet{kultima2015developers}, who note that developers regard iteration as an essential and natural part of the development process. However, we identified three primary approaches to design decision-making: \textbf{(1)} an \emph{Academically-Grounded, Theory- and Research-Driven Approach}, which draws systematically on formal models and empirical evidence; \textbf{(2)} an \emph{Experience-Based Approach}, which is systematic but shaped by industry practices and practical constraints; and \textbf{(3)} a \emph{Gut Feeling-Driven Approach}, characterized by non-systematic, individual preferences informed by personal experience and intuition. 
\\

\textbf{\textcolor{fierypurp}{\faChevronCircleRight}~\ul{Academically-Grounded, Theory- and Research-Driven Approach}}
\\
Experts emphasized that when using academic research, it is critical to incorporate those insights early in \textbf{pre-production}, while the project is still taking shape. As {\color{burntorange}\faUser~Expert 5} noted, ``You need those types of information [research] at the start to guide quick decisions in the future. Because at that point [later in production] you don’t have the time to do it.'' Similarly, {\color{crimsonred}\faUser~Expert 4} reflected that in pre-production, ``You have the most time to explore [...] once it’s done, priority goes elsewhere, and getting a mandate for change is really hard.''

    \displayquoting{When we are in conception of the game, [I] look if papers have been released or published on the topics that we will work on.}
    {{\color{black}\faUser~Expert 15}}

From our semi-structured interviews, our experts often referred to established frameworks (and design workflows). The frameworks raised most frequently were Nielsen's Heuristics~\cite{nielsen1994}, Gestalt Principles\footnote{\href{https://www.interaction-design.org/literature/topics/gestalt-principles?srsltid=AfmBOoow7oQqrbbRzetG3YrRy-b1cdv4xW7jG9RnL2p-fM4vMqrFwHXa}{Gestalt Principles} describe how people perceive visual elements as grouped wholes, e.g., by proximity, similarity, closure, or continuity. They originate in Gestalt psychology and are widely used in visual and interface design.}, rapid A/B testing~\cite{quin2024b}, the Design Space Model~\cite{Fagerholt_Lorentzon_DesignSpaceModel_2009}, and UX laws such as Fitts' Law~\cite{fitts1954information}, Murphy's Law, Miller's Law~\cite{miller1956magical}, Tufte's Principles~\cite{tufte1990}, and Atomic Design\footnote{\href{https://atomicdesign.bradfrost.com/table-of-contents/}{Atomic Design} is an interface design methodology by Brad Frost that structures design systems into five hierarchical levels: atoms, molecules, organisms, templates, and pages, building interfaces from small reusable components.}, often referred to under acronyms or alternative labels. Cognitive Load Theory and several cognitive biases (e.g., Peak-End Rule, Serial Position Effect) were also commonly mentioned, though by fewer experts. Frameworks such as Self-Determination Theory~\cite{deci2012self}, MDA~\cite{hunicke2004mda}, Accessible Player Experience Guidelines~\cite{beeston2018accessible}, and game metaphors were raised by only a few, while some reported using no specific framework at all. 

Rather than aggregating responses, we organized them by how commonly each framework was mentioned, which surfaced patterns in familiarity, perceived relevance, and integration within AAA studio practices.\footnote{Following our research ethics, we focus on frameworks used in practice rather than experts' disciplinary backgrounds.} Several experts drew on a broader ``bag of knowledge'' ({\color{teal}\faUser~Expert 10}) from cognitive science, interaction design, and related fields without naming specific frameworks. Interestingly, experts did not attribute their academia-inspired work to the actual, formal theories. Instead, our experts had a tendency to ``develop their own internal narratives and classifications and names'' ({\color{teal}\faUser~Expert 10}), with the core goal of developing a ``shared understanding, a language'' ({{\color{plum}\faUser~Expert 2}}) to align as a team and have the same language to describe things. One expert explained how they shape Nielsen's heuristics:

    \displayquoting{I don't call them heuristics because it's not important for anyone to know. I translate them into ``ambitions''. When I say we want to be able to accomplish tasks quickly, we want to have sufficient signs and feedback all the time. And we want consistent behaviours when it comes to selection and confirming and cancelling [...]. It's all about standardization and managing expectations and terminology.}
    {{\color{navyblue}\faUser~Expert 12}}

    \displayquoting{I use some stuff from academia, but it's about how can we apply them in practice. You know, how can I use this to make people look at the game they're creating, not at the theory?}
    {{\color{plum}\faUser~Expert 2}}
    
Experts highlighted that frameworks often neglect the commercial and organizational realities of game development, further limiting practical adoption as: ``[i]n academia, you focus on the users, [i]n industry, we also have to focus on business needs and team constraints'' ({\color{emeraldgreen}\faUser~Expert 9}).

    \displayquoting{I think in practice it's often hard to apply everything because you are already working in a space that is very confined, especially working on a big game where you don't like to build the system from scratch.}
    {{\color{crimsonred}\faUser~Expert 4}}

While experts acknowledged the value of research-driven theory, they emphasized significant challenges to its practical application. Academic approaches were often described as misaligned with industry realities—too abstract, time-consuming, or disconnected from business priorities. The distinctive complexity of games further limited their applicability. As one expert noted, there is a ``mismatch between established theory and the actual user experience'' ({\color{olive}\faUser~Expert 7}). Moreover, most UX frameworks originate in web design, offering limited guidance tailored to the unique demands of game development.

    \displayquoting{UX and UI frameworks [...] mostly relate to web interfaces, not so much to the game realm. I would love for the frameworks we know and love to be understood [and] applied in a game design context. [...] we occupy a very unique space that has a lot of crossover with others, but [...] decisions are made at a creative level, not necessarily on a product level.}
    {{\color{goldenrod}\faUser~Expert 6}}

    \displayquoting{Models are great [...] but I still have to do the work of making them relevant for me.}
    {{\color{navyblue}\faUser~Expert 12}}

In addition, several experts observed that scientific terminology is sometimes referenced without being meaningfully applied. Such terminology was often perceived as alienating to colleagues without research backgrounds, and reading papers under time pressure was described as hard work with little payoff. Even theories that seemed relevant were still not ``100 percent transferable to production'' ({\color{indigo}\faUser~Expert 1}).

    \displayquoting{I can talk about self-determination theory in my design, but actually never do a design that caters to that [...] they are mainly using it for selling their work better off [...] So that's gonna end up as a failure.}
    {{\color{plum}\faUser~Expert 2}}

These patterns suggest that theory informed awareness and early-stage thinking, however, practitioners frequently reinterpreted and adapted academic concepts before integrating them into production. 
\\

\textbf{\textcolor{fierypurp}{\faChevronCircleRight}~\ul{Experience-Based Approach (Systematic with an Industry Lens)}}
\\
Many practitioners described approaches that were not directly based on the literature (though they were sometimes inspired by it) but were primarily built on accumulated professional experience, previous projects, and team-specific needs. These approaches were not full-fledged processes or pipelines, but rather often took the form of internal guidelines, style guides, ``interaction bibles,'' or design pillars---documents that codified best practices for the project while aligning cross-disciplinary teams. 

    \displayquoting{We design the general rule set for how the UI is gonna work. Then, maybe another team is working on a little feature prototype, and they want to build a little interface for it, right? So we try to provide them with a documentation saying like if you build an interface for the keyboard and mouse, then it should be connected like this, and these are the rules that you should take into consideration.}
    {{\color{forestgreen}\faUser~Expert 8}}

Games are not only sophisticated interactive systems but also unique, creative products that balance functionality with entertainment, novelty, and immersion. 
    
    \displayquoting{The gaming industry works on products that are incredibly challenging, perhaps the most challenging interactive products out there.}{{\color{teal}\faUser~Expert 10}}
    
Designers must consider hundreds of interconnected interfaces, from trade routes to character interactions. They must also adapt designs across various platforms, including consoles, PCs, and mobile devices, each with distinct guidelines.

    \displayquoting{Due to the intentional lack of control we exert on the players, a game is seen as a very different product, with unmatched interactivity and player agency.}{{\color{forestgreen}\faUser~Expert 8}}
    
Many experts rejected comparisons between games and other digital products, arguing that universal frameworks rarely work without customization. Moreover, many saw their own games as unique even within the industry, requiring bespoke solutions.

    \displayquoting{None of this logic existed before us [...] it has to be something special}
    {{\color{forestgreen}\faUser~Expert 8}}

Two convictions, in particular, reinforced this reliance on experience over formal theory: that games are too distinctive for off-the-shelf frameworks, and that genuine competence in the field is built experientially rather than learned from models. Experts framed formal frameworks as optional rather than foundational, pointing to accomplished colleagues whose practice rested entirely on accumulated craft knowledge.

    \displayquoting{I see a lot of great UX professionals who can do their job fine without them.}
    {{\color{plum}\faUser~Expert 2}}

    \displayquoting{People go by their proven backgrounds. It's like their own experience, their knowledge.}
    {{\color{indigo}\faUser~Expert 1}}

This experiential orientation is partly rooted in the composition of the workforce itself. Game designers come from highly diverse backgrounds, and there is no standardized educational pathway into game UX. Some have only completed brief boot camps with limited exposure to formal frameworks, while others transition from unrelated fields such as architecture, sociology, or production, bringing craft sensibilities from their prior disciplines rather than shared theoretical foundations.

    \displayquoting{I’ve seen motion designers become UX designers.. I’ve seen baristas become UX designers... we’re all a mishmash.}
    {{\color{navyblue}\faUser~Expert 12}}

As a result, UX decision-making in these contexts becomes highly person-dependent, with individual preferences and informal power dynamics frequently shaping design choices alongside---or in place of---structured, research-informed approaches.
\\

\textbf{\textcolor{fierypurp}{\faChevronCircleRight}~\ul{Gut Feeling-Driven Approach (Non-Systematic Individual Preferences with Personal Lens)}}
\\
A limited group of experts---though notable given the AAA scale of their studios---reported working without any formalized UX processes in pre-production. 
{\color{deepblue}\faUser~Expert 11}, for example, highlighted that their team has deliberately agreed upon a fully informal approach that relies on ad-hoc discussions and consensus-building to make design decisions, without following any specific frameworks or models. 

In these cases, decision-making relies on designers’ individual expertise, instincts---a gut feeling---and interpersonal influence within the studio. {{\color{raspberry}\faUser~Expert 3}} alluded to designers' ``more intrinsic understanding of things'', while {{\color{forestgreen}\faUser~Expert 8}} argued that ``a senior [designer] will work on instinct, they will instantly see that something works or doesn’t.'' Internal politics and informal discussions often shaped design sign-offs:

    \displayquoting{I think you just have to convince the key people, talk to them. I think there's no magic solution for this. [...] I happen to be really good friends with the game director. A lot of late night beers, basically, and arguing in bars.}
    {{\color{raspberry}\faUser~Expert 3}}

This reliance on personal judgment was partly linked to diverse educational backgrounds. Some designers entered the field without formal UX training, making shared frameworks less central to their practice.

    \displayquoting{That's where experience comes in. If you have a lot of formal training in UI/UX theory, then you will probably apply those things, and if you come from a different background that's not something that you consider to be very important.}
    {{\color{forestgreen}\faUser~Expert 8}}

Many experts indicated that academic work is underused, often because designers are unaware of these publications or do not know how to incorporate the frameworks into their design processes. However, several participants described working with academic concepts in an approximate, informal way. They acted on a ``hunch'' about a framework's validity without being certain they were applying it faithfully or using its terminology correctly. 

Rather than a binary of use versus non-use, academic knowledge appears to enter practice in loose fragments. These fragments are useful enough to act on but too vague to share across teams or defend in design discussions.

    \displayquoting{I just don’t know of them [...] because I don’t know them, I don’t know if I’m actively using them or not.}
    {{\color{olive}\faUser~Expert 7}}

    \displayquoting{I think in a lot of these cases, it's I'm trying to find, I have a hunch about the validity. [...] I don't know if I'm using the exact correct framework. I'm not using the terms correctly. But I know this to be a thing, and therefore this is how I'm dealing with that situation. I'd rather we do that than nothing.}{{\color{goldenrod}\faUser~Expert 6}}

Experts also cautioned that partial gut feeling knowledge could be counterproductive. Without proper training, frameworks risk being misapplied. 

    \displayquoting{If you think you understood something but didn’t, it can give you a sense of security and false results.}
    {{\color{burntorange}\faUser~Expert 5}}
    
Thus, while formal education can help spread awareness, it also highlights the need for maturity and a shared understanding across teams. However, many experts emphasized that academic works still hold important value for the industry.

\subsection{\texorpdfstring{\textcolor{fieryred}{\faCircle}}~Theme 2: Designers Operate in Transversal, Collaborative Roles Across the Production Pipeline to Align Design with Player Needs, Technical Constraints, and Creative Vision}\label{theme_2}

Designers in AAA studios did not work as isolated specialists but in transversal roles that cut across the production pipeline, sustaining collaboration with art, engineering, narrative, production, and business functions from the earliest stages to align design with player needs, technical constraints, and creative vision.

    \displayquoting{I work not just with us designers, but with artists, developers, engineers and tech designers. Everything to make sure that the direction we set is both feasible, but also that everyone is kind of bought in.}
    {{\color{deepblue}\faUser~Expert 11}}

Designers described their work as ``touching base'' with multiple departments---often three or four at a time ({\color{indigo}\faUser~Expert 1})---navigating organizational complexity that includes ``hierarchy, spread-out teams, different types of producers, and the monetization department'' ({\color{crimsonred}\faUser~Expert 4}).

Experts introduced two dominant collaborative structures used to manage complexity and distribute workload in large-scale game projects: \textbf{strike teams} and \textbf{competency teams}. These structures enable both rapid, interdisciplinary feature development and sustained, expertise-driven support across the project lifecycle.
\\

\textbf{\textcolor{fieryred}{\faChevronCircleRight}~\ul{Strike Teams (Cross-Functional, Feature-Focused Units)}}
\\
\textbf{Strike teams}---sometimes referred to as \emph{cells}---are small, cross-functional groups formed to rapidly explore, design, and implement specific features. These teams typically include members from game design, UX- and UI design, programming, and quality assurance (QA)\footnote{From our reflexive perspective, we highlight the distinction between quality assurance (QA) and quality control (QC) from our experts. QA refers to a process-oriented, proactive approach that aims to prevent defects by embedding quality throughout the development or production process. In contrast, QC is product-oriented and reactive, focusing on detecting defects through testing and inspection of the final product before it is released.}. In cases where features are more narrative-driven, narrative and gameplay designers may also be integrated. A producer commonly oversees the strike team, ensuring that the work remains within scope and is delivered on schedule.

    \displayquoting{UI is like a touch point to a lot of the other neighboring departments. So basically, these are small strike teams which are cross-functional. So let's say, for instance, one of the team members works on the mini map. Then there will be one member from game design, one UI designer, UI programmers or gameplay programmers, and someone from QA inside. If we have a feature that is more story-related, we will also have gameplay designers and narrative designers inside [...] Most of these features are then run by a producer, who makes sure that these features are delivered in scope and on time.}
    {{\color{indigo}\faUser~Expert 1}}

These interdisciplinary units support tightly coupled collaboration by placing designers, coders, UX specialists, and QA professionals within a coordinated communication structure.

    \displayquoting{We call these [interdisciplinary teams] ``cells''. A cell is a team and every cell has 1---2 or multiple features that are connected to the cell. [...] It's designers, it's coders, it's UX, it's QA. It's everybody under one place, and it's a very well-structured communication structure.}
    {{\color{black}\faUser~Expert 15}}

The breadth of these collaborations is itself telling: strike teams drew not only on core craft disciplines but also on business, data, and legal expertise, underscoring how far designers' transversal role reaches beyond the creative pipeline into the wider organization. Some practitioners even highlighted close collaborations with experts such as data analysts, monetization experts, or legal experts, if relevant for the respective feature development and its ideation.

    \displayquoting{There's the monetization department; so for any feature touching monetization or any adjacent kind of rewards, there's an entire own team for that.} {{\color{crimsonred}\faUser~Expert 4}}

    \displayquoting{[When there's] a more ecommerce side of the product, I had to work a lot with the legal team. (Features with) money and payment require a lot of legal regulations, depending on the region.}
    {{\color{emeraldgreen}\faUser~Expert 9}}
    
    \displayquoting{I do talk a lot with our data analysts, particularly on the last feature I did [...], because it was supposed to be a price history graph. [So] I need to have this understanding, and it needs to work for both perspectives.}
    {{\color{olive}\faUser~Expert 7}}

While strike teams streamline feature development through intense collaboration, experts noted that their delivery focus can leave limited time for broader exploration or research review.
\\

\textbf{\textcolor{fieryred}{\faChevronCircleRight}~\ul{Competency Teams (Expertise-Driven, Discipline-Based Clusters)}}
\\
In contrast to the feature-oriented nature of strike teams, \textbf{competency teams} are organized around areas of specialized knowledge (e.g., accessibility, animation, or technical UI implementation). These individuals are considered subject-matter experts and often serve as the \emph{first point of contact} for domain-specific guidance across the wider organization.

    \displayquoting{In addition to the generalist approach, some people specialize in certain fields, like accessibility or animation, and these team members are considered [...] subject matter experts.}
    {{\color{indigo}\faUser~Expert 1}}

Rather than owning individual features, competency teams provide consultation, establish best practices, and help maintain consistency and quality across multiple strike teams and project components. These teams commonly consist of a combination of highly specialized roles, including UI technical artists, UX designers, UI programmers, and UI artists, and typically report to a subject-matter lead who oversees various divergent areas of a broadly shared expertise (in comparison to the strike teams that are usually coordinated by producers):

    \displayquoting{We call that the ``presentation team''. UI art plus UX. We also had a UI tech artist, and they were tech-expert enough to discuss directly with the programmers. They were doing reviews with them and were able to say what is feasible and what is not.}
    {{\color{darkgray}\faUser~Expert 14}}

Within these teams, experts emphasized that the least amount of friction is not always beneficial. Instead of pursuing effortless interactions as a blanket principle, teams collectively examined where friction may support emotional impact, player engagement, or narrative meaning. These discussions often occurred during cross-disciplinary reviews, where team members compared perspectives and ensured that decisions about friction aligned with both usability expectations and the project's broader creative direction.

    \displayquoting{So it's not always the quickest, least friction to path. Sometimes friction is a good thing. It depends on the context. But if it was just about completing a task or job to be done, like - I have managed my character, I have equipped the best item on there - it loses the emotional impact of why you did it and how it made you feel empowered or smarter, or that you've grown over time or created shorter long-term goals.}
    {{\color{goldenrod}\faUser~Expert 6}}

While not all AAA game productions use the exact terminology of strike teams and competency teams, most experts adopted a similar complementary system when resources allow, combining interdisciplinary units to support both rapid feature development and sustained expertise. In combination with strike teams, competency teams create a dual-layered structure that balances speed, creativity, and deep domain expertise within complex game development workflows.

\subsection{\texorpdfstring{\textcolor{fieryoj}{\faCircle}}~Theme 3: Stakeholder Priorities and Awareness Gaps Are Key Barriers to Adopting Frameworks in Complex Game Projects}\label{theme_3}

Experts also elaborated on academia as a tool for stakeholder presentations, reinforcing design decisions ``because if they were already leaning towards accepting your design [...] it gives them extra validation that science supports (this) as well'' ({\color{forestgreen}\faUser~Expert 8}). However, while references to research-driven theory can sometimes support conversations with senior stakeholders, the majority of experts questioned their overall usefulness in day-to-day practice.

    \displayquoting{I don't think it's that useful [...] If I pull theory, it can help in some cases, unless you know the theory. (If not), am I going to spend half an hour explaining Fitts's law to someone?}
    {{\color{slate}\faUser~Expert 13}} 
    
For most internal discussions, explicitly invoking design principles that stakeholders are unfamiliar with was seen as unhelpful and, in some cases, even counterproductive. If theoretical language is not shared across the group, it can effectively ``stop the discussion'' ({\color{plum}\faUser~Expert 2}).

    \displayquoting{Trying to justify things [with] design principles, but stakeholders are not necessarily aware.}
    {{\color{forestgreen}\faUser~Expert 8}}

Experts described how design decisions are shaped not only by player needs but also by the competing demands of technical feasibility, production schedules, business priorities, and monetization. As {\color{emeraldgreen}\faUser~Expert 9} explained, designers must integrate perspectives from ``customers, teammates, managers, and stakeholders'' into a coherent outcome.

    \displayquoting{We always do reviews with the monetization team [...] to make sure they are happy.}
    {{\color{olive}\faUser~Expert 7}}

Stakeholders often impose priorities that outweigh theoretical considerations; in some cases, rules for interaction emerge reactively from production realities rather than being defined upfront through research. Under these conditions, designers often lack the autonomy to apply academic works, even when they desire to do so.
\\

\textbf{\textcolor{fieryoj}{\faChevronCircleRight}~\ul{Pre-Production as the Critical Window and the Impact of Production Demands on Design Decisions}}

Experts described pre-production as the only phase where there is space, albeit still limited, to engage with research, exploration, or frameworks; once development advances, deadlines and emergencies dominate.

    \displayquoting{The best value of my professional expertise will be at the beginning.}
    {{\color{teal}\faUser~Expert 10}}
    
    \displayquoting{There's always the point in the preparation phase where you have more time to look at [research]. Then, at some point in the production, you know the train is rolling [...] Production timelines I wouldn't say completely rule them out, but they narrow the scope.}
    {{\color{indigo}\faUser~Expert 1}} 

Several experts noted that production pressure is not about disinterest but about survival: ``Everyone is incredibly busy and stressed in game development'' ({\color{deepblue}\faUser~Expert 11}). Under these conditions, even useful guidelines are often abandoned in favor of quick fixes.

    \displayquoting{Sometimes it’s like hey, tomorrow is the deadline [...] I need a solution that will fix that thing in three hours, and a guideline will not do that.}{{\color{burntorange}\faUser~Expert 5}}

Experts emphasized that stakeholder priorities create an environment that encourages ``bare minimum'' solutions, reactive firefighting, and trade-offs between quality and feasibility. They explained that designers are constantly engaged in cost–benefit discussions, in which they decide whether to implement a quick but inelegant fix or to pursue a more thoughtful, player-centric design that ideally should have been addressed during pre-production rather than in the middle of production.

    \displayquoting{I went like [slur] about how we need to do as an MVP because I felt like I was getting way too much work and I was on too many teams and overworked. So I was very strict about [doing] the bare minimum [they] needed me to do so.}
    {{\color{olive}\faUser~Expert 7}}

Experts across studios consistently identified pre-production as the key phase for exploration and design decision-making.   
\\

\textbf{\textcolor{fieryoj}{\faChevronCircleRight}~\ul{Research Lacks Visibility and Actionable Formats for Industry Application}}
\\
While the practitioners expressed genuine interest in frameworks, they consistently described research as difficult to access and hard to understand, which forced them to rely on informal, practice-oriented sources. The resulting disconnect reveals a twofold knowledge-transfer problem. First, there is a challenge of \textit{discoverability}, which is knowing which studies or frameworks exist and how to locate them. Second, there is a challenge of \textit{translatability}, which involves adapting formal theories into actionable, context-sensitive practices suitable for game development. One expert, reflecting on their own transition from academia to industry, was shocked at how little academic work was directly applicable or immediately usable in day-to-day production decisions.

    \displayquoting{Myth and rubbish is prevalent and you have consultants who come in and give courses about things they don't know anything about.}
    {{\color{slate}\faUser~Expert 13}}

Because formal research was often difficult to access and apply, experts turned to more immediate and socially embedded channels. Social media platforms (e.g., LinkedIn, Twitter/X, YouTube, BlueSky), private peer networks, and industry-focused conferences became their primary sources of inspiration and knowledge. 

    \displayquoting{I like conference talks because it's coming through the lens of someone who's taken experience and time and application.}
    {{\color{navyblue}\faUser~Expert 12}}

Practitioners described these more immediate and socially embedded channels as preferable because they present distilled insights that are already contextualized for game development.
\\

\textbf{\textcolor{fieryoj}{\faChevronCircleRight}~\ul{Experts Call for Concrete, Contextualized Use Cases}}
\\
When we asked our experts how academic work could be more useful, they offered a set of requirements. They emphasized the need to visualize principles: abstract principles are challenging to convey across teams, whereas visual representations facilitate immediate understanding.

    \displayquoting{Visualizing what works and what doesn’t [...] like simple charts or graphs that I could give to a junior or another designer in another department.}{{\color{forestgreen}\faUser~Expert 8}}
    
Our experts stressed their need for concrete use cases, which are grounded in the realities of game development. They noted that the generalized principles discussed in academic work were perceived as less valuable than tangible and applicable examples, particularly on online platforms such as Medium, UX Collective, or Nielsen Norman Group. 

    \displayquoting{Use cases, always. That's by far the most tangible thing.}
    {{\color{deepblue}\faUser~Expert 11}} 

They praised online resources like the Game UI Database\footnote{\href{https://www.gameuidatabase.com/}{The Game UI Database} is a free, comprehensive online collection of user interface (UI) elements, images, and videos from games on many platforms and genres. For game UI and UX designers, it is a preferred reference that offers thorough classification and search filters to help them discover ideas, look up specific UI implementations, and see how industry peers have addressed design problems.} because they provide exactly this kind of applied inspiration. Experts also highlighted a clear need for guidance on multi-platform development. With most AAA titles shipping across PCs, consoles, and other devices, designers often lack systematic methods for adapting interfaces across different input methods and hardware. They called for ``very pragmatic, concrete things'' ({\color{deepblue}\faUser~Expert 11}) to ensure consistent user experiences across platforms.  

Across interviews, experts consistently described the disconnect as stemming from the difficulty of accessing and applying research in production contexts rather than from a rejection of academic rigor. Designers expressed a demand for theory presented in formats aligned with fast-paced, collaborative workflows---visual, grounded in concrete use cases, and adapted to multi-platform challenges.

\subsection{\texorpdfstring{\textcolor{fieryblue}{\faCircle}}~Theme 4: Shared Language and Common Ground Can Enable Stronger Collaboration Between Academia and Practice}\label{theme_4}
Beyond individual design tasks, experts saw frameworks as a tool to strengthen collaboration and create a shared language within teams. In large, complex projects, successful design depends on coordinating work and aligning perspectives across disciplines. If appropriately adapted, frameworks could facilitate such collective understanding, serving as a common point of reference across teams and stakeholders. This highlights that game development is as much about people as it is about products.

    \displayquoting{Design has to bring people together. [...] Focus on the people is my starting and ending point [...] on one end, the player, but also the people that work on the game themselves.}
    {{\color{teal}\faUser~Expert 10}} 

Collaboration across diverse expertise---designers, engineers, artists, producers---was described as both essential and difficult. {\color{raspberry}\faUser~Expert 3} elaborated that while design principles are useful, the greater challenge often lies in ``actually solving group dynamics and working together.'' In this sense, frameworks or guidelines were not valued solely for their prescriptive content, but for how they could help teams establish a shared vocabulary and coordinate effectively.

Experts noted that frameworks and shared vocabularies could support collaboration at scale---helping interdisciplinary teams establish common reference points and reduce misunderstandings that often slow production (as noted in \hyperref[theme_2]{Theme 2}). While \hyperref[theme_1]{Theme 1} explored how designers adapt terminology for cross-disciplinary communication, here we examine the role of language and collaboration in fostering the adoption of academic works.
\\

\textbf{\textcolor{fieryblue}{\faChevronCircleRight}~\ul{Design Systems as the Foundation for Shared Vocabularies and Understanding}}
\\
Experts highlighted shared practices that academia could leverage, particularly \textit{design systems}, which ``solved a lot of problems'' ({\color{navyblue}\faUser~Expert 12}) and improved collaboration between designers and other experts ({{\color{crimsonred}\faUser~Expert 4}}). Experts explained that design systems are typically established during pre-production by a small, dedicated team and then extensively used throughout production by designers and developers, helping mitigate the impact of time constraints and tight deadlines. {\color{darkgray}\faUser~Expert 14} stated, ``We created the process and the guidelines around the design system,'' hinting at how the design system shaped other aspects of their design and process. Other experts also noted how design systems streamlined design and development activities:

    \displayquoting{It's a very easy thing to do because we have everything in the design system. And I say everything, even the distances, like when you create a new UI element, it's pretty much decided where it will go and how it will be.}
    {{\color{black}\faUser~Expert 15}}
 
Experts also indicated that integrating design systems into the workflow reduces the time of exploration and simplifies the problem-solving approach. However, when a design system is not established during pre-production, its absence can become a bottleneck later in the development process.

    \displayquoting{I don't need to think too much about where to put things in the UI and how big they are, because it's already established for us by the global UX community and [...] also in our design system. We basically have a whole typography design system, element sizes and console support.}
    {{\color{black}\faUser~Expert 15}} 

    \displayquoting{It is a problem because often we are working within a lot of constraints, because there was no alignment in structure. There are certain elements of the UI that we cannot change. That's something that we learn very, very early, like global components that are always present, like that's just something that we cannot touch unless there's a bigger mandate for it or like a redesign mandate}{{\color{crimsonred}\faUser~Expert 4}}

Similarly, {\color{indigo}\faUser~Expert 1} emphasized how extensive design systems can be, as they can include ``design patterns, elements, [...] buttons, sliders, and [all kinds of] functions and elements.'' The need for shared language was echoed by {\color{plum}\faUser~Expert 2}, who envisioned a sort of extension of design systems---an \textit{artifact} that collects all of a game’s content, including features, systems, and definitions, into one accessible place, enabling everyone to ``look at it holistically and talk about the game.'' Academic works could play a role here by offering not only theories but also concepts that teams can adopt and integrate into their internal terminologies. As {\color{teal}\faUser~Expert 10} suggested, research could help teams extend their vocabulary with ``science-inspired design'' concepts that bridge external knowledge and internal ways of working.

Experts described design systems as a foundation for building shared vocabulary and understanding, and suggested that research presented in compatible formats could integrate more readily into existing or future design systems.
\\

\textbf{\textcolor{fieryblue}{\faChevronCircleRight}~\ul{``Conceptual Lego Pieces'' as a Basis for Assembling Custom, Modular Frameworks}}
\\
At the same time, experts were clear that such frameworks must remain flexible, like individual pieces. Rigid, ``by-the-book'' applications of academic works were seen as ill-suited to the dynamic and creative realities of game production. 

    \displayquoting{Games are such individual pieces of software that it is often hard to work with generalized information. [...] (Frameworks should) be considered less of a prescription and more like the guidelines-approach of showing common issues, limitations [...] and a direction.}
    {{\color{crimsonred}\faUser~Expert 4}}

    \displayquoting{So rules are only important until they are not, and especially on the design side, we have to be kind of open for this. We have to make sure that the UI/UX is not trying to force something onto the game [...] there's a lot of systems which work together in different combinations to create this new scenario which might need a unique visualization or feedback-solution to reinforce that kind of uniqueness.}
    {{\color{forestgreen}\faUser~Expert 8}}

Practitioners noted that most frameworks they consider somewhat useful were not specifically developed for games ({{\color{indigo}\faUser~Expert 1}}) and therefore asked instead for modular resources that could be tailored to each project’s needs.

    \displayquoting{[U]sing bits and pieces of whatever you need for your own model framework that pulls all the attention to the product, the game.}
    {{\color{plum}\faUser~Expert 2}}

{\color{teal}\faUser~Expert 10} described these as ``conceptual Lego pieces'' — discrete heuristics, ontologies, and guidelines that designers could assemble into context-specific frameworks. This approach allows teams to engage with academic contributions selectively, while also mitigating the risk of misinterpretation or inappropriate application by practitioners who may lack formal academic training or the full context of the original research.

    \displayquoting{A space of possibility to create a collection of concept ontologies, heuristics and guidelines that all together can help designers think in a broader way [...] and bring in some good arguments about measurable [...] this is where the design credibility comes from.}
    {{\color{teal}\faUser~Expert 10}}

In addition, every designer has their own unique style, highlighting the importance of collaborative decision-making to ensure these modular pieces remain flexible and the final outcome is shaped collectively.

    \displayquoting{I've seen a lot of the different designers who come up with the UI design, and it's like their food, their styles were all different. Personally, my style is I do come up with a lot of designs like good or bad. I explored around 9 different versions of it, and I think we should decide with the team what they like, what I like, and then discuss it.}
    {{\color{emeraldgreen}\faUser~Expert 9}}

These perspectives collectively suggest that practitioners valued academic contributions primarily as resources for collaborative sense-making and for building what experts called ``design credibility'', grounding decisions in shared, recognizable reasoning.

%% file: sections/04_Disc.tex
\section{Discussion}\label{discuss}

The \hyperref[results]{Results} section highlighted persistent gaps and the complex realities of design practice, particularly during \textbf{pre-production}. Building on this, our work extends existing models of knowledge translation in HCI. \citet{colusso2019translational} map the continuum largely from the vantage point of researchers and translators who carry knowledge toward practice; we examine the same TAD gap from the practitioner side. Our contribution is therefore not a new gap but an empirical account of practitioner-led adaptation, and of stakeholder governance as the structure constraining it. 

\begin{figure}[!ht]
    \centering
    \includegraphics[width=1\linewidth, trim={1cm 3cm 5cm 2.7cm}, clip]{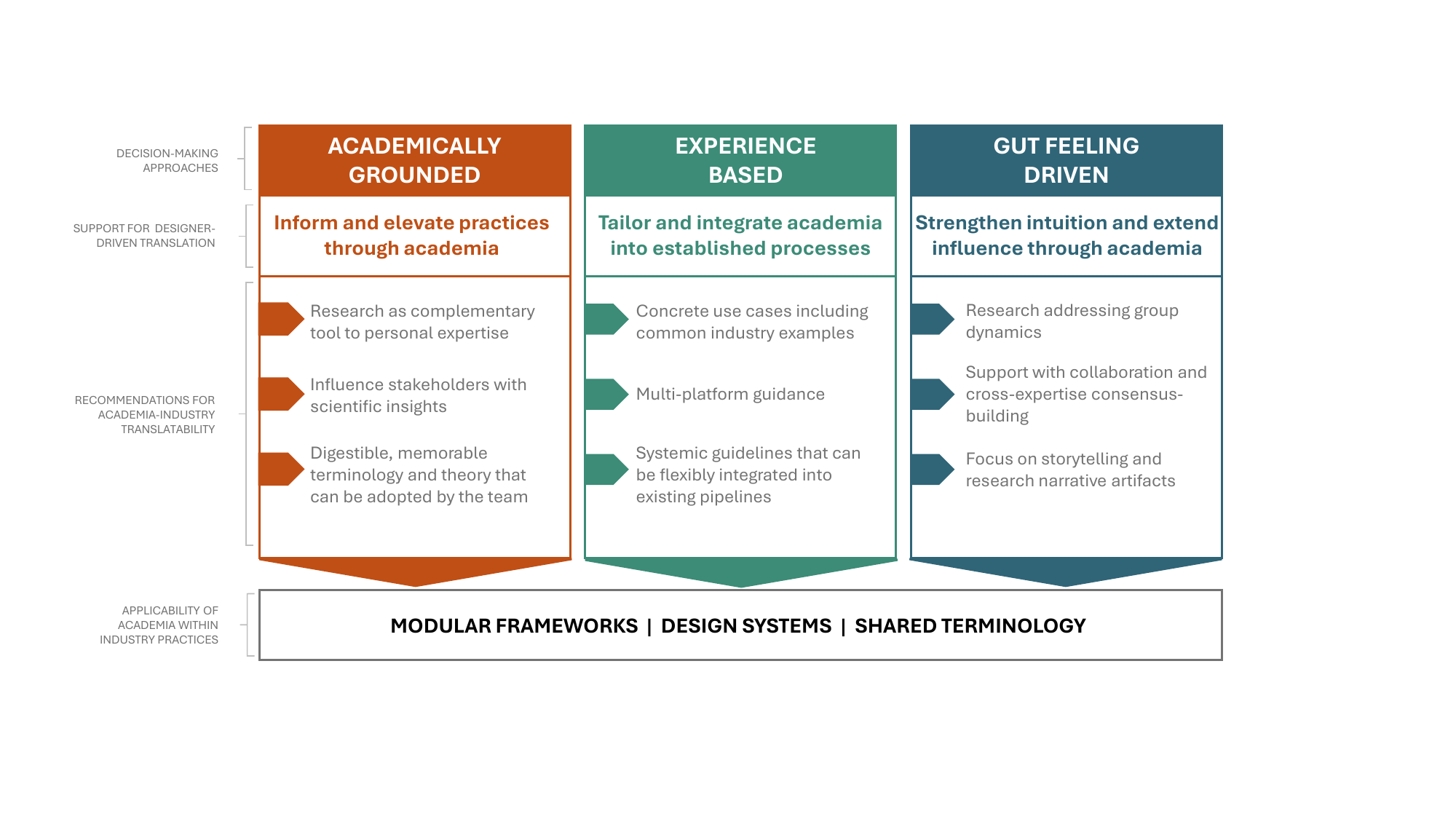}
    \caption{\textbf{Applicability of academic deliverables within established industry practices.} Specific recommendations on how to tailor the translatability of academia for practitioners, based on different decision-making preferences. While projects, teams or company culture might nurture one or more of the three approaches to creative design decision-making, there is no single ``right'' method; designers move fluidly among approaches depending on context.}
    \Description{This figure illustrates strategies for translatability of research within an industry context. Three pillars describe decision-making approaches of academically grounded, experience-based, and gut feeling grounded, along with recommendations for specific designer-driven translation and applicability of research within industry practices.}
    \label{fig:ADJ_model}
\end{figure}

Our findings suggest that bridging research and practice requires recognizing the limits of academic prescription, consistent with previous work~\cite{o2019game,kultima2015developers,karabinus2018games, randelshofer2026theory}. The \textit{how} of design is contingent on studio culture, structure, and project constraints. Academic contributions are therefore most valuable when they clarify the \textit{what} by offering high-level, empirically grounded principles that practitioners can adapt to their own contexts. In the remainder of this section, we first introduce the Model of Adaptive Design Judgment (RQ1). We then examine the transversal team structures---strike teams and competency teams---and the coordination constraints that determine where and how academic knowledge can enter AAA practice (RQ2). Finally, we derive implications for how academic research can more effectively support large-scale game development practice (RQ3).

\definecolor{experience}{HTML}{26826B}
\definecolor{gutfeeling}{HTML}{326578}
\definecolor{academic}{HTML}{D99781}

\subsection{{Approaches to Early-Stage Design Ideation and Creation in AAA Game Development (RQ1)}}\label{RQ1}

In this section, we describe three overlapping approaches to design decision-making: academically grounded (practitioner-led translation of academic theory), experience-based (codification of tacit knowledge), and gut feeling driven (intuitive, informal processes). Practitioners draw on these fluidly and in combination during early-stage design. Synthesizing our findings across multiple AAA studios, we integrate them into a descriptive model of Adaptive Design Judgment (ADJ) that characterises how practitioners describe navigating complex, high-pressure design contexts. As a synthesis of practitioner accounts rather than a tested construct, the model is intended as an analytic lens and a starting point for empirical validation rather than an evaluated framework.

While some teams deliberately avoided processes (\hyperref[theme_2]{\textcolor{fieryred}{\faCircle}~Theme 2}), the majority---including UX leaders who relied on gut feeling approaches---valued established processes that incorporated academic works. The lack of formal academic works (\hyperref[theme_3]{\textcolor{fieryblue}{\faCircle}~Theme 3}) often reflected company cultures that favored individual stakeholder decisions over formalized practices. In sum, these findings map onto the three pillars of the ADJ model in AAA game development (\autoref{fig:ADJ_model}), which pairs each pillar with recommendations for designer-led translation and for enhancing the applicability of research within industry practice. By framing our findings within this model, we provide a \textit{coherent lens} for understanding real-world design practices.
\\

\textbf{\ul{Practitioner-Led Translation of Academia (Academically Grounded)}}

Our findings indicate that practitioners draw inspiration from design frameworks but rarely apply them in their original textbook form. While \citet{o2019game} argue that no single framework is universally ``best'' and must instead be right-fit for different game types, our findings suggest that adoption varies not only by genre but also by studio structure, production scale, leadership strategy, and stakeholder influence. Practitioners do not apply theory passively; they actively \textbf{reinterpret it}. They extract core concepts selectively and recast them as localized shared languages (\hyperref[theme_1]{\textcolor{fierypurp}{\faCircle}~Theme 1} and \hyperref[theme_2]{\textcolor{fieryred}{\faCircle}~Theme 2}), turning formal heuristics into constructs that carry internal meaning for the team (e.g., ``ambitions''). Because these adaptations are designer-led, they often become detached from their original theoretical grounding and are reshaped around project-specific constraints, particularly where sustained \textbf{researcher–practitioner–stakeholder} collaboration is absent (\hyperref[theme_3]{\textcolor{fieryblue}{\faCircle}~Theme 3}). This pattern reflects the long-standing HCI research–practice gap~\cite{Colusso2017Translational, Velt2020}.

Importantly, our findings suggest that the research–practice gap does not disappear at the level of UX leadership. Even UX leaders, who are positioned to translate academic insight into strategy, operate within stakeholder-driven structures where commercial imperatives, risk management, and executive priorities shape what can be implemented. This points to stakeholder governance as a central structural factor shaping the gap. Studio scale further amplifies this dynamic. In smaller studios, flatter structures allow novel methods to circulate more directly~\cite{colby2019game}, even if collaboration remains informal~\cite{mcdaniel2016developer, ruggill2016inside}. In contrast, AAA studios rely heavily on established, industry-validated artifacts such as competitor benchmarks (\hyperref[theme_1]{\textcolor{fierypurp}{\faCircle}~Theme 1}) to coordinate across specialized teams (later discussed in \autoref{RQ2}). These shared reference points reduce interpretive overhead but also constrain the integration of emerging frameworks. Under milestone-driven schedules and multi-stakeholder dependencies, the flexibility required to meaningfully adapt academic work is significantly reduced~\cite{politowski2021game} (\hyperref[theme_3]{\textcolor{fieryblue}{\faCircle}~Theme 3}, \hyperref[theme_4]{\textcolor{fieryoj}{\faCircle}~Theme 4}). 

\begin{callout}
Therefore, drawing on the positionality of UX leaders, we argue that research should be designed with greater sensitivity to stakeholder governance structures, industry timelines---particularly pre-production phases---and production constraints. Only then can frameworks function as effective boundary objects that meaningfully bridge academia and AAA development practice.
\end{callout} 


\textbf{\ul{Codifying Tacit Knowledge for Experience and Design Systems (Experience Based)}}

Practitioners also grounded their approach in accumulated professional experience and team-specific practices. While practitioners did not formally attribute these approaches to specific frameworks (e.g., playtesting frameworks), the approaches were often informed by such frameworks, which studios had embedded in their culture. Teams valued such processes for their pragmatism and adaptability, yet their transferability remained uncertain. As discussed in (\hyperref[theme_1]{\textcolor{fierypurp}{\faCircle}~Theme 1}), designers primarily build expertise through situated, experience-based practice rather than through formalized theoretical models. This aligns with \citet{Sturdee2022}'s research on situated practices, showing that expertise is built experientially and resists codification into universal frameworks. Previous work further shows that designers routinely address ill-defined problems through cycles of reframing and solution development \cite{micheli2019doing}, and that expert practice relies heavily on tacit knowledge, rapid pattern recognition, and skilled intuition \cite{schindler2015expertise, wong2000tacit}. 

Our findings extend this literature by showing that this experience-based approach does not remain purely tacit. Practitioners actively work to externalize and stabilize their knowledge through shareable artifacts such as \textit{interaction bibles} and \textit{design systems} (\hyperref[theme_1]{\textcolor{fierypurp}{\faCircle}~Theme 1}, \hyperref[theme_4]{\textcolor{fieryoj}{\faCircle}~Theme 4}). Rather than merely relying on intuition, they engage in an ongoing process of making experiential knowledge explicit and scalable within teams. This codification process constitutes a significant, yet often overlooked, form of design labor. Hence, we argue that academics should support and study these \textbf{codification practices as legitimate design activities}, contributing principles and methods that integrate directly into internal knowledge systems rather than over-generalizing their own approaches. \\

\textbf{\ul{Gut Feeling, Informal Processes, and Team Mechanisms (Gut Feeling Driven)}}
\\
In some game studios, design decisions are often guided by gut feeling, informal processes, and the influence of visionary leaders\footnote{For example, Hideo Kojima, Shigeru Miyamoto, and Swen Vincke are widely recognized in the game industry as visionary leaders, known for shaping the creative direction of projects such as the Metal Gear series, Nintendo franchises like Mario and Zelda, and Baldur's Gate 3 and the Divinity series, respectively, regardless of the commercial or critical reception of individual games.}, exemplifying what is often referred to as an \textit{auteur} culture. Some practitioners deliberately maintain ad-hoc processes, making decisions through discussions, consensus, and experienced designers' instincts rather than formal frameworks. Participants described relying on gut feeling and interpersonal influence within the resources available at AAA studios (\hyperref[theme_1]{\textcolor{fierypurp}{\faCircle}~Theme 1}). Although HCI research identifies intuition and social dynamics as central to collaborative design \cite{feng2023collab, schaffer_study_2018}, gut feeling also emerges as a form of expert judgment. Senior practitioners navigate substantial uncertainty and use intuition as an effective decision‑making tool, which aligns with theories of expert practice such as~\citet{schon_reflective_2017}’s concept of the reflective practitioner. The success of these non-systematic teams also depends on mechanisms beyond individual intuition, including consensus-building discussions, interpersonal influence, shared mental models, and relational trust among team members (\hyperref[theme_2]{\textcolor{fieryred}{\faCircle}~Theme 2} and \hyperref[theme_4]{\textcolor{fieryoj}{\faCircle}~Theme 4}). These mechanisms allow teams to collectively adapt to uncertainty even without formalized processes. However, these non-systematic teams implicitly enact principles aligned with Design Thinking (DT). They approach projects as exploratory spaces, treating problems as open-ended and iterative---similar to how cultural probes are used in research~\cite{gaver1999design}. Just as probes provide practitioners with materials to generate unexpected insights and reflections rather than definitive answers, practitioners of gut feeling use informal discussions and iterative feedback to gather inspiration and guide their decisions. 

By engaging in reflective, collaborative, and iterative problem-solving, which mirrors DT methods such as ethnographic observation for needs finding, brainstorming for idea generation, and storytelling or prototyping to visualize solutions, they navigate pre-production uncertainty and complex stakeholder demands~\cite{seidel2013adopting, lockwood2009transition, cooper2009design}. DT here functions both as a reflective mindset and a structured approach to the fuzzy front end of product development, helping teams create value and align around a shared vision, even in the absence of formalized frameworks as suggested by~\citet{de2021acquaintances}. 

\begin{callout}
Thus, we argue that academia can support these informal structures by providing \textbf{shared vocabulary, and team-level mechanisms for alignment, negotiation and consensus}, enhancing the collective effectiveness of non-systematic teams while acknowledging the role of auteur culture in shaping creative decision-making.
\end{callout}

\subsubsection{Reflexive Standpoint}

We ask ourselves the question ``so what'' then, given our positionality in academia and industry. In terms of academic work, our finding means that researchers need to reconsider how they design theories, methods, and contributions if they hope to influence practice. If translation fails not because ideas are unclear but because organizational structures distribute authority, risk, and ethical responsibility in uneven ways, then academic outputs must be developed with these conditions in mind. This requires shifting academic assumptions about impact~\cite{madathil2019succeed,sotamaa2021introduction, politowski2021game}. In contrast to the view that translation is primarily a matter of dissemination or clarity, our results indicate that it succeeds or fails within organizational settings shaped by commercial pressures, stakeholder priorities, and competing ethical commitments. Understanding translation as a structural and political problem directs future research to the parts of industry practice where academic interventions can matter most.

\subsection{Transversal Roles and Constraints in AAA UX Design Aligning Design with Player Needs, Technical Constraints, and Creative Vision (RQ2)}\label{RQ2}

Having identified stakeholder governance as the structural root of the translation gap (\autoref{RQ1}), we now examine the team structures and production decision-making through which it operates in AAA studios. Our analysis indicates that two transversal team structures, \textit{strike teams} and \textit{competency teams}, jointly shape how academic knowledge is translated into AAA practice by coordinating short-horizon feature work with longer-horizon standards and expertise. This pairing determines where academic interventions can be integrated, how quickly they must operate, and which forms of evidence or artifacts are most likely to gain traction.

To understand where academic work can better support industry practice, we focus here on these two team structures that play central roles in game development. \hyperref[theme_2]{\textcolor{fieryred}{\faCircle}~Theme 2} indicates that smaller \textit{strike teams}, composed of cross-functional members from art, design, programming, narrative design, and other areas, are effective for rapidly developing specific features and managing organizational complexity. Strike teams bring together members from these disciplines to quickly explore and deliver a specific feature or gameplay mechanic, often working closely with other departments due to the intertwined nature of game development. In parallel, \textit{competency teams}, which focus on specialized areas such as UX design, accessibility, and technical implementation, provide domain-specific expertise and act as the first point of contact when challenges arise. \\

\textbf{\ul{A Walkthrough of Decision Making in Practice}}
\begin{figure}[!ht]
    \centering 
    \includegraphics[width=1\linewidth, trim={0cm 0cm 0cm 0cm}, clip]{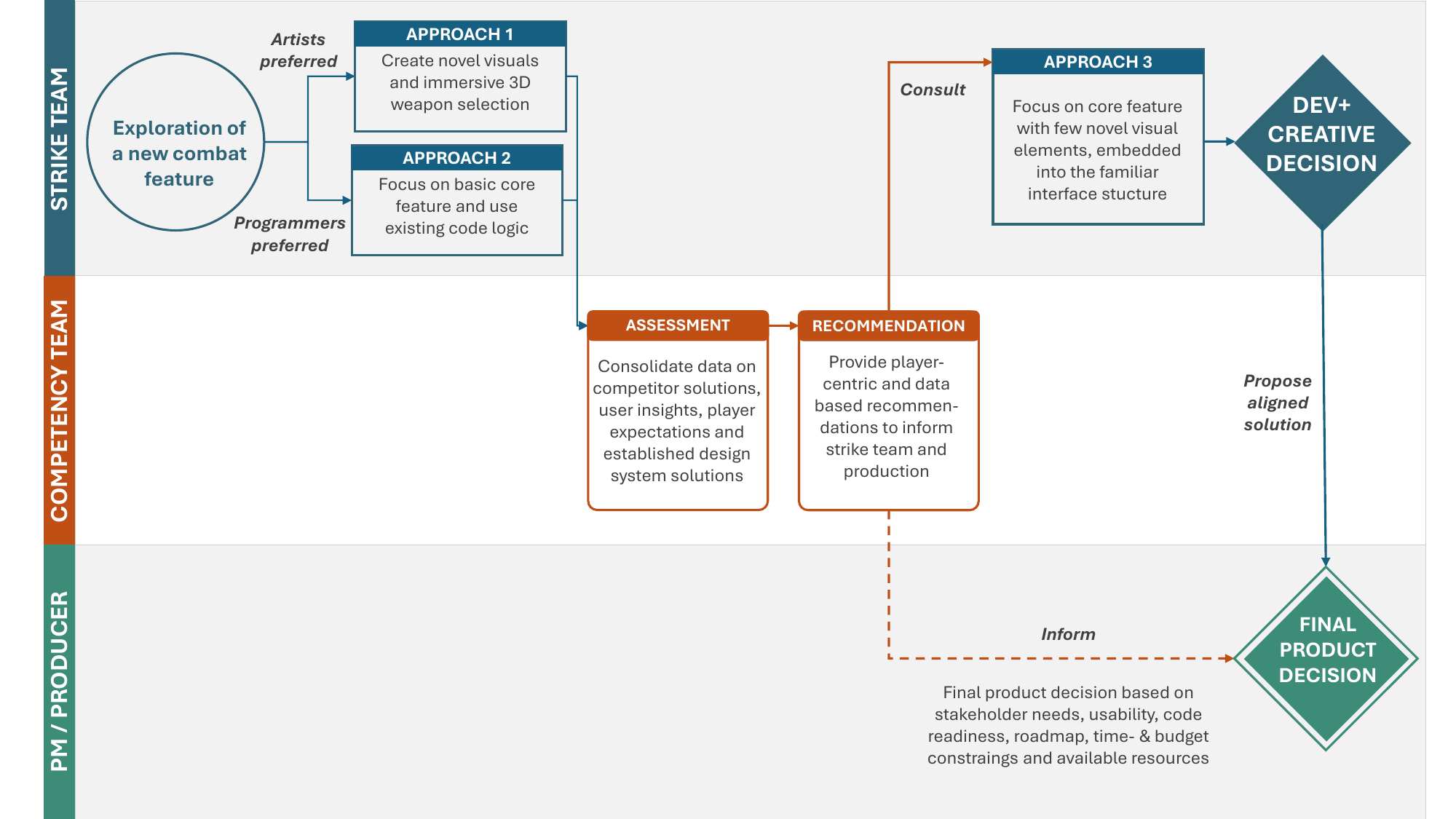}
    \caption{An illustrative example to demonstrate the decision-making process between strike team, competency team, and production management (PM) in developing new features.}
    \Description{This presents a flowchart that illustrates the close collaboration among a strike team, a competence team, and production during the exploration of a new combat feature. The strike team ideates on different approaches and trade-offs between artist-preferred and programmer-preferred solutions. The competence team informs and consults their next step by consolidating data, user insights, accessibility constraints and design system recommendations. The final decision about the feature approach is made by production after assessing stakeholder needs, time, and budget constraints and available resources.}
    \label{fig:striketeam_flow}
\end{figure}

For example, a strike team ideating on a new combat mechanic might encounter conflict in balancing technical constraints with feature usability. Artists may prefer creating novel visuals and immersive 3D weapon selection, while programmers may focus on the core feature and reuse of existing code. These alternative approaches are then passed to the UX competency team, which assesses them using data on competitor solutions, user insights, and player expectations. The team provides a recommendation that balances both perspectives---such as focusing on the core feature while embedding a few novel visual elements---thereby consulting and informing the strike team before an aligned solution is proposed. Such a process allows development to continue without delays, as seen in \autoref{fig:striketeam_flow}. This layered approach helps mitigate bottlenecks, streamline communication across the production pipeline, and balance agility with strategic oversight. Operating simultaneously at these two scales enables studios to enhance innovation and decision-making, extending previous work on interdisciplinary integration with professional UX practitioners \cite{feng2023collab}.

In AAA contexts, strike and competency teams reflect principles of co-design methodologies by enabling interdisciplinary collaboration and knowledge exchange. Work by \citet{steen2013co} describes co-design as an iterative process that integrates multiple stakeholder perspectives throughout development. For instance, \citet{bekius2023framework} present a framework to involve stakeholders in the design process using sequential, looped learning (adapting \citet{bloom1994reflections}'s taxonomy): stakeholders must reach \textit{Understand} before \textit{Apply} phases, with subsequent levels depending on prior learning. However, achieving higher levels is constrained by factors such as time, problem urgency, available data, participant expertise, and group dynamics.

\definecolor{strike}{HTML}{2596be}
\definecolor{comp}{HTML}{d99781}
\definecolor{pm}{HTML}{3e8c77}

\begin{table}[ht!]
\centering
\small

\begin{tabular}{@{}l | p{8.5cm}@{}}
\toprule
\textbf{Proposed Practice} & \textbf{Suggested Framework Adaptation} \\ 
\midrule
\rowcolor{strike!10}~\textcolor{strike}{\faClock}~Time-boxed artifacts & Concise, stepwise guides that fit within sprint or milestone boundaries, with explicit estimates for effort and dependencies. These can be quickly adopted by \textbf{strike teams} while accounting for perspectives from artists, programmers, and other contributors. \\[1mm]
\midrule
\rowcolor{comp!10}~\textcolor{comp}{\faClipboardList}~Competency-ready standards & Consolidated assessments from the \textbf{competency team}, including competitor analysis, user insights, and established design systems, to provide player-centric guidance that informs \textbf{strike teams} and production workflows. \\[1mm]
\midrule
\rowcolor{pm!10}~\textcolor{pm}{\faChartBar}~Evidence formats that travel & Findings translated into comparable benchmarks, simple scorecards, and canonical examples that can be referenced across teams without extensive interpretation, ensuring clarity for Production Managers (PM), producers and other stakeholders. \\[1mm]
\midrule
\rowcolor{fierypurp!10}~\textcolor{fierypurp}{\faSync}~Alignment to production phases & Recommendations mapped to pre-production, vertical slice, and content-lock stages, highlighting when adoption offers the highest leverage and lowest risk. \\
\bottomrule
\end{tabular}
\caption{Proposed adaptations for embedding academic research within existing AAA development workflows. These adaptations are proposals grounded in practitioner accounts and remain to be empirically evaluated.}
\Description{Table 4 is a two‑column, four‑row table titled ``Actionable practices and suggestions for framework adaptation and modification for existing AAA game development workflows and established frameworks.'' The first column lists four actionable practices, each preceded by a small colored icon. The second column provides a detailed suggestion for each practice. Row one shows a blue circle icon with the label ``Time‑boxed artifacts,'' paired with a description explaining that teams should use concise, stepwise guides that fit within sprint or milestone limits, including effort estimates and dependencies, and suitable for cross‑disciplinary strike teams. Row two shows an orange square icon with the label ``Competency‑ready standards,'' paired with a description stating that consolidated assessments such as competitor analysis, user insights, and design systems should provide player‑centric guidance for strike teams and production workflows. Row three shows a green square icon with the label ``Evidence formats that travel,'' paired with a description recommending that findings be translated into benchmarks, scorecards, and canonical examples that can be easily shared and understood across teams, including producers and PMs. Row four shows a red square icon with the label ``Alignment to production phases,'' paired with a description advising that recommendations be mapped to pre‑production, vertical slice, and content‑lock stages to indicate when adoption offers the most benefit and least risk.}
\label{tab:req}
\end{table}

Rather than introducing new frameworks (and workflows) to the Games User Research (GUR) and HCI communities, we recommend strengthening existing ones by incorporating the modification practices and addressing the challenges identified in \hyperref[theme_4]{\textcolor{fieryoj}{\faCircle} Theme 4}. Accordingly, we propose adaptations that, based on participants' accounts, could make research more actionable within AAA workflows, as summarized in~\autoref{tab:req}. These adaptations are derived from how practitioners described their existing workflows rather than from any evaluation of the adaptations themselves; we therefore offer them as proposals for future empirical validation rather than as demonstrated solutions.

Research has shown that agile development promotes iterative cycles, cross-functional collaboration, and responsiveness to change~\cite{al2020agile,cao2009framework}. However, AAA production operates at a much larger scale, with highly specialized teams, complex asset pipelines, and milestone-driven dependencies that constrain flexibility once features enter implementation.

For example, Accessible Player Experiences (APX) promote actionable design guidance~\cite{beeston2018accessible}, yet operationalizing these recommendations in AAA contexts requires navigating diverse accessibility needs alongside technical integration and production constraints. The core challenge, therefore, lies in disconnects across production phases, terminology, and levels of complexity within AAA studios. As highlighted in \hyperref[theme_3]{\textcolor{fieryblue}{\faCircle} Theme 3}, frameworks must not only serve end users but also align with business objectives and team constraints~\cite{Petrillo2009}.

\begin{callout}
Thus, we suggest that future frameworks and guidelines address key challenges in AAA game development, including \textbf{structural constraints of rapid iteration and decision-making processes between strike teams, competency teams, and production management}, and the need to \textbf{account for coordination across multiple teams and stakeholders}, so that practitioners can adopt, adapt, and extend them to fit their projects and team structures.
\end{callout}

\subsection{Applicability of Academic Works In Practice for Industry Use (RQ3)}

Rather than reiterating the Model of Adaptive Design Judgment, we focus here on its implication. While integrated knowledge translation (IKT) models emphasize sustained co-production~\cite{Graham2018MovingKnowledge}, our findings suggest that such depth of collaboration is structurally limited in large-scale game production. Instead, research must be prepared for constrained environments characterized by distributed teams, production dependencies, and layered stakeholder governance~\cite{Fidas2015, colusso2019translational}. Our findings suggest that, to gain traction in these settings, translation-ready knowledge would need to be modular, recombinable, and compatible with existing coordination mechanisms and production stages (as suggested in ~\autoref{tab:req}).

    \begin{table}[!htbp]
    \centering
    \small
    \begin{tabular*}{\columnwidth}{@{\extracolsep{\fill}}p{0.2\columnwidth}p{0.35\columnwidth}p{0.35\columnwidth}@{}}
    \toprule
    \textbf{Strategy} & \textbf{Suggested Implementation} & \textbf{Anticipated Impact} \\ 
    \midrule
    Design System & Create structured rules, reusable components, and guidelines for consistent UX, UI and gameplay & Provides shared reference, facilitates collaboration, supports onboarding \\ 
    \midrule
    Documentation & Document design rules, technical requirements, and team practices clearly & Maintains project alignment, accelerates framework adoption, supports distributed teams \\ 
    \midrule
    Concrete Use Cases & Apply academic works to specific game components & Improves applicability, enables direct testing, demonstrates reusability \\ 
    \midrule
    Participatory Approach & Incorporate user feedback iteratively throughout design & Builds trust, improves experience, aligns with company and user needs \\ 
    \midrule
    Socio-Technical Alignment & Integrate design, technical requirements, and collaboration using STS perspectives & Fosters cross-departmental work, enhances framework adoption, addresses organizational dynamics \\ 
    \midrule
    Knowledge Dissemination & Use workshops, conferences, and internal sharing instead of journals & Reduces gut feeling reliance, increases research awareness, accelerates adoption \\ 
    \bottomrule
    \end{tabular*}
    \caption{\small Six strategies for enhancing framework applicability in design practice. Suggested implementation methods and anticipated impacts.}
    \Description{Table 5 provides six strategies that can be used to enhance the applicability of academic frameworks in design practice. These strategies include the use of design systems, the importance of documentation, the presentation of concrete use cases, the application of participatory approaches, the alignment of socio-technical considerations, and the dissemination of knowledge. Each of these strategies is paired with a description of its implementation and the impact it can have on design outcomes.}
    \label{tab:acad-strategies}
    \end{table}  

Contrary to assumptions that frameworks struggle to gain traction in AAA contexts, \hyperref[theme_4]{\textcolor{fieryoj}{\faCircle} Theme 4} highlights the central role of design systems in coordinating game development work. Our findings suggest that design systems function as socio-technical infrastructures that mediate between design, code development, accessibility, and team collaboration, aligning with previous work~\cite{Isbister2018social}. As \citet{daSilva2016} argues, design systems shape social behaviour and nurture cross-departmental collaboration by embedding organizational dynamics into structured practices. In this sense, they act as a ``North Star'' guiding designers, developers, marketers, and content writers through shared standards and reusable components~\cite{Naseva2024}. Beyond visual consistency, they unify gameplay mechanics, interaction patterns, accessibility standards, and brand identity into a single source of truth~\cite{Oppermann2022}.

However, while design systems provide stability and coordination, our findings also reveal a tension: their emphasis on standardization, particularly of visual components, can risk narrowing interpretive flexibility. As reflected in \hyperref[theme_3]{\textcolor{fieryblue}{\faCircle} Theme 3}, an overreliance on predefined components may obscure holistic player experience considerations and, in some cases, constrain innovation or weaken the player journey.

As mentioned in \autoref{tab:req}, having findings translated across multiple teams is important: it drives consistent practice, efficient onboarding, and cross-team alignment with project values and culture~\cite{megan_doc_2011, mcdonald2021describing}. Design systems support this translation in concrete ways. For example, a design system might specify that all health indicators use red, all interactive elements respond within 200 milliseconds, and all tutorial prompts appear in the upper-left corner. Such rules create predictable patterns that players can learn quickly, and---in the process---they bridge gut feeling and frameworks by making tacit design decisions explicit and transferable. While these strategies may seem intuitive, \autoref{tab:acad-strategies} clarifies how they function as coordinated interventions, linking specific implementation mechanisms to tangible impacts that support adaptation, cross-team alignment, and sustained integration of research within production workflows. Below, drawing on publicly available industry examples beyond our interview data, we illustrate how these strategies are currently used in practice.

\begin{figure}[ht!]
    \centering
    \includegraphics[width=1\linewidth]{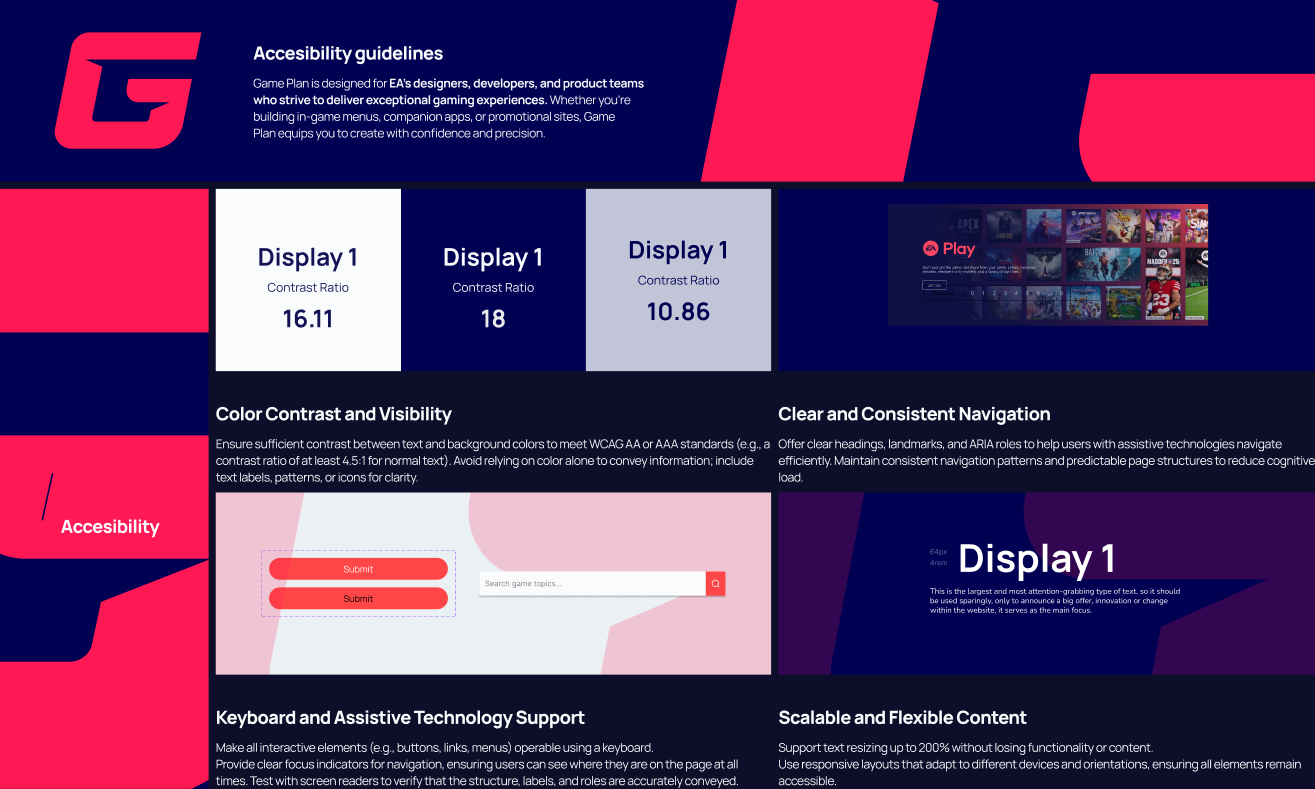}
    \caption{Accessibility principles from EA Games' design system for menu interfaces.}
    \Description{This figure illustrates how accessibility principles are integrated into menu creation at EA Games. The example shown is drawn directly from EA Games’ design system known as the GamePlan, and it demonstrates how accessibility guidelines are embedded within the design process.}
    \label{fig:acc-design}
\end{figure}

\textbf{\ul{Design System ``Currently'' in Practice}}\label{design_system}

EA Games' recently published design system\footnote{EA's Design System (2025) called \href{https://www.figma.com/community/file/1443366636816575662}{GamePlan} provides a unified foundation for seamless collaboration across teams.} illustrates how accessibility principles guide menu creation. When designing menus for \textit{The Sims}\footnote{\textit{The Sims} is a life simulation video game series developed by Maxis and published by Electronic Arts.}, these guidelines standardize layout, typography, color contrast, and navigation to support all players (see \autoref{fig:acc-design}). Implementing these strategies under production pressures in AAA development is challenging (\hyperref[theme_3]{\textcolor{fieryblue}{\faCircle}}~Theme 3). Iterative user feedback requires rapid testing protocols, recruiting representative users, scheduling sessions, and analyzing results promptly to inform design---processes that are time- and resource-intensive~\cite{halskov2015diversity, spinuzzi2005methodology}. Similarly, applying a socio‑technical systems (STS) lens involves structured cross‑department workshops, alignment meetings, and integration of social, organizational, and technical requirements across teams---overhead that often discourages full adoption~\cite{baxter2011socio, mirri2018collaborative}. While these strategies can improve collaboration, framework adoption, and design quality, their feasibility under tight deadlines depends on careful scoping, prioritization, and selective application, or on building meta‑design infrastructure that supports ongoing participation rather than one‑off exercises.
\\

\textbf{\ul{Translational Work Theory ``Currently'' in Practice}}

\hyperref[theme_2]{\textcolor{fieryred}{\faCircle}~Theme 2} and \hyperref[theme_4]{\textcolor{fieryoj}{\faCircle}~Theme 4} emphasize that frameworks need concrete use cases and reusability. Despite these potential benefits of design systems, our experts reported that they relied on peer learning through conferences, workshops, and books when challenges arose for them (\hyperref[theme_3]{\textcolor{fieryblue}{\faCircle}~Theme 3}). This preference aligns with \citet{mcdaniel2016developer}'s findings that game developers favour informal technical communication over formal documentation. Academic works stayed mostly neglected because journal articles (and peer review turnaround times) do not match industry's faster, practice-oriented knowledge exchange~\cite{Rogers2004}. \citet{Norman2010} refers to this as a \textit{research-practice gap}, where findings take years to reach practitioners, while \citet{Perkmann2013AcademicEngagement} notes that even when academics engage with industry, knowledge transfer remains limited to informal relationships rather than published works. \citet{Graham2018MovingKnowledge} calls this the \textit{know-do gap}, where valuable research fails to translate into actionable practice. 

\begin{figure}[ht!]
    \centering
    \includegraphics[width=1\linewidth]{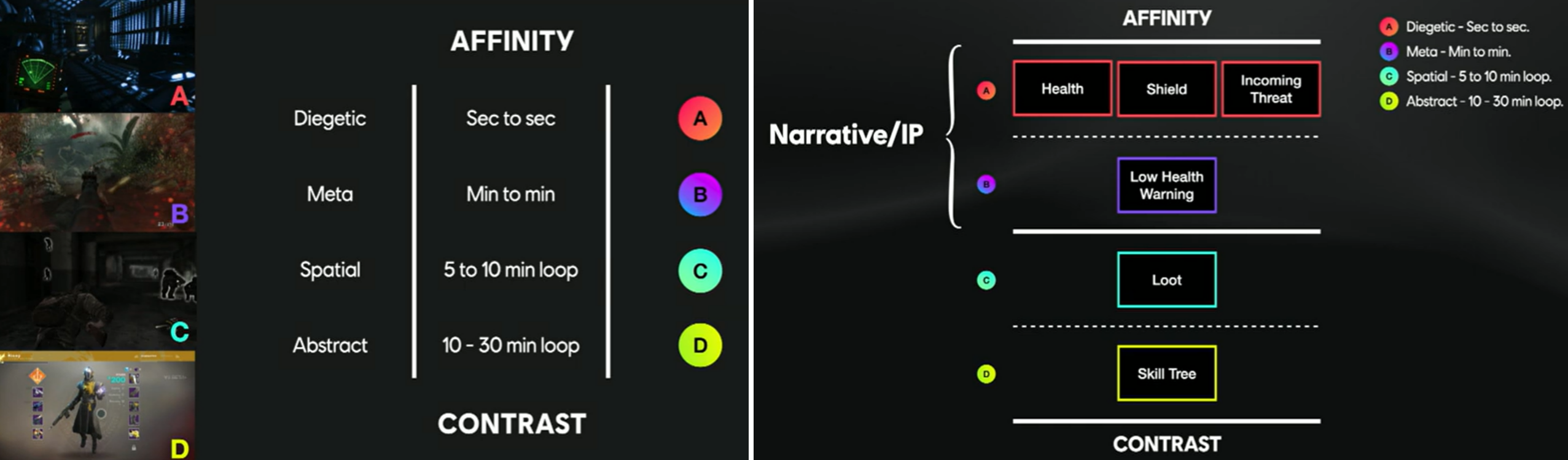}
    \caption{Salama categorizes interface features by type, interaction frequency, and duration to guide designers in grouping and visualizing interfaces for player attention and engagement. Diegetic elements, such as health information, remain constantly in focus, whereas abstract interfaces, like HUDs, menus, and skill trees, are accessed in 10–30 minute intervals. He suggests that diegetic and meta displays, which are inherently narrative-driven, may benefit from collaboration with the storytelling or narrative department---an input that might otherwise be overlooked in the design process.}
    \Description{This figure categorizes interface features by type, interaction frequency, and duration to help designers group and visualize interfaces in relation to player attention and engagement. The left side presents an affinity-based classification with four descriptors—diegetic, meta, spatial, and abstract—while the right side breaks down the narrative/IP into sub-levels of specificity: diegetic elements include health, shield, and incoming threat; meta elements feature low health warnings; spatial elements encompass loot; and abstract elements include the skill tree.}
    \label{fig:salama_presenceaffinity}
\end{figure}

A concrete and applicable example for designer-driven translation work in practice can be found in the case study presented by Ahmed Salama (former UX Content Director from Ubisoft). In a masterclass\footnote{Presented at \href{https://conference.digitaldragons.pl/speakers/ahmed-salama/}{Digital Dragons conference} and Devcom 2024 workshops. As these conferences and workshops are not formally citable, permission to use these slides was obtained from Ahmed Salama.}, he introduced a framework that re-purposed academic frameworks to fit production needs, guiding decision-making during development. Salama leverages established terminology such as~\textit{diegetic interfaces, HUDs, meta displays, spatial representations} from~\citet{Fagerholt_Lorentzon_DesignSpaceModel_2009} to player perception and presence---a mapping he developed from his industry experience" or "...presence, drawing on a theory developed from his industry experience. This ``conceptual Lego'' approach mixes academic and practical knowledge, creating diagrams that estimate player attention across interfaces as suggested in {\textcolor{fieryoj}{\faCircle}~Theme 4}. In this work, he also strengthens cross-disciplinary collaboration, indicating which experts should be consulted for specific interfaces. Salama highlights that combining theory and industry experience serves as a guiding tool shared across UX teams, game designers, and other departments to establish a cohesive vision, common terminology, and support holistic reasoning about the game, as illustrated in \autoref{fig:salama_presenceaffinity}. These contributions demonstrate how practitioners effectively utilize structured, theory-informed tools to inform real design decisions and maintain a coherent UX vision throughout the development process.

%% file: sections/05_Conclusion.tex
\section{Limitations and Future Work}
Our findings should be read alongside the boundaries of our study design, each of which points to specific future work. Our sample comprises exclusively senior, lead-level practitioners at AAA studios, and is predominantly male; the findings therefore capture a leadership vantage point within one segment of the industry. Future studies should interview junior and mid-level UX designers---ideally in leader--report dyads within the same studio---to test whether the adaptation work we describe is visible to, shared by, or contradicted by those executing the decisions. Comparative work in indie, AA, and mobile contexts, alongside purposive sampling of women and non-binary UX leaders, would further establish which of our themes are AAA-specific and which reflect leadership demographics rather than the practice itself.

Our data are also retrospective self-reports, subject to recall bias and post-hoc rationalization, and individual-level accounts cannot characterize studio-wide processes. Observational methods---workplace ethnography, diary studies during active pre-production, or think-aloud protocols in design reviews---would reveal whether theory and framework adaptation operate in situ as practitioners describe them, while multi-informant case studies of single studios, triangulating leaders, reports, and internal documentation, could describe decision-making at the organizational level. Finally, we measured neither design outcomes nor framework use, so our recommendations remain directions for evaluation rather than validated interventions. A natural next step is a longitudinal deployment study: co-designing modular translation artifacts with one or more studios, then tracking their adoption, adaptation, and perceived influence on pre-production decisions over a project cycle.

\section{Conclusion}
Our study examined how UX leaders in AAA game studios make pre-production design decisions, revealing that decision-making arises from a blend of theory, experience, and intuition---an adaptive, context-sensitive practice that formal knowledge-translation models rarely capture. We believe academic theories have a firm place in the world of UX leaders, but they must be designed to support the adaptive realities of practice: translated into shared language, embedded in collaborative artifacts, and delivered as modular frameworks. Our work contributes a descriptive model of this reality, complementing prescriptive models of knowledge translation by unearthing the crucial, often-invisible work of practitioner-led adaptation that occurs when formal integration is not possible. We intend this model as a description of how leaders reason about integration, not as evidence that any particular adaptation strategy improves design outcomes. In the end, our task is not to dictate to designers \textit{how} to do their practice, but to enrich \textit{what} their thinking in their practice is about.

%% file: sections/06_Appendix.tex
\newpage
\section{Semi-Structured Interview Questions}\label{app:interview_questions} 

We provide an unordered list of the semi-structured interviews. Questions are mainly focused towards pre-production phase. To protect participant anonymity and comply with Non-Disclosure Agreement (NDA) requirements and institutional ethical guidelines, any questions related to company-specific processes, such as titles in ideation or active development, were excluded from the data synthesis due to the potential professional consequences for participants, including risk to employment (e.g., designers disagreeing with stakeholders’ monetary decisions or conceptual game development within the pre-production phase). As many of our experts relied on \textit{industry slang} and insider terminology, we adopt a reflexive position to clarify how these expressions were interpreted within the context of our analysis. \\

\textbf{Collaboration and Workflow}
\begin{itemize}
    \item Can you tell us about your role in the game development process?
        \begin{itemize}
            \item Where do your contributions have the biggest impact and/or least impact?
        \end{itemize}
    \item Who do you collaborate with most frequently?
        \begin{itemize}
            \item E.g., UX Designers, Researchers, Game designers, Developers, Managers, Audio Designers, Accessibility Specialists, QA testers
        \end{itemize}
\end{itemize}

\textbf{Design Process}
\begin{itemize}
    \item Can you please walk us through your typical process when creating X (eg; when creating user interfaces)?
    \item Which stage of the development process are you most involved?
    \item Who else contributes to this design process?
        \begin{itemize}
            \item How do you hand off your work for further development?
        \end{itemize}
    \item Do you and/or your team have a formal or informal process when designing interfaces, especially in the early pre-production stages?
        \begin{itemize}
            \item Do you follow a strict pipeline, and/or is there more room for improvisation?
        \end{itemize}
    \item Are there established guidelines you follow during design?
    \item Do you encounter any challenges during your design work on game interfaces?
        \begin{itemize}
            \item If so, what are they, and how can they be addressed?
        \end{itemize}
\end{itemize}

\textbf{Design Principles and Frameworks}
\begin{itemize}
    \item Do you rely on any established design principles, frameworks, or guidelines during design and interface development? (Focus on creation, not evaluation methods)
    \item (If they mention principles/frameworks) Can you elaborate on one you find particularly effective?
        \begin{itemize}
            \item Where did you learn about it, and how do you apply it?
        \end{itemize}
    \item Have you used these principles to advocate for your designs in presentations to stakeholders?
        \begin{itemize}
            \item Was the reception rather positive, negative or neutral?
        \end{itemize}

    \item Which aspects of these frameworks are most valuable and efficient for the design expertise?
        \begin{itemize}
            \item (e.g., guidelines, visuals, terminology, use cases)
        \end{itemize}
    \item Are there any drawbacks to relying on formal *guidelines/theories/terminology*? (see above)
    \item What additional resources could support better design decision-making, especially in the pre-production phase?
    \item Hypothetically speaking, if we were to create a framework for your team and project, what would be the most important deliverables of such a framework, so it would be usable, applicable and beneficial for you?
        \begin{itemize}
            \item E.g. concrete use cases, focus on games context, visuals, patterns, scientific grounding, theories, and etc.
        \end{itemize}
\end{itemize} 
\thispagestyle{empty}

\textbf{Continuous Learning}
\begin{itemize}
	\item Are there any other design frameworks or resources you would like to mention?
	\item How do you stay informed about the latest trends in game design?
    \begin{itemize}
        \item E.g., online articles, conferences, podcasts, academic research and etc.
    \end{itemize}
	\item What format do you prefer for learning about new design principles and frameworks?
\end{itemize} 

\textbf{Closing Questions}
\begin{itemize}
	\item Do you have any questions for us?
	\item Is there anything else that you would like to mention regarding this topic?
\end{itemize}

\section{Participant Quotes Mapped to the Model of Adaptive Design Judgment}\label{app:theme_mapping} 

\begin{figure}[h!]
    \centering
    \includegraphics[width=0.5\textheight, height=0.6\textwidth, keepaspectratio]{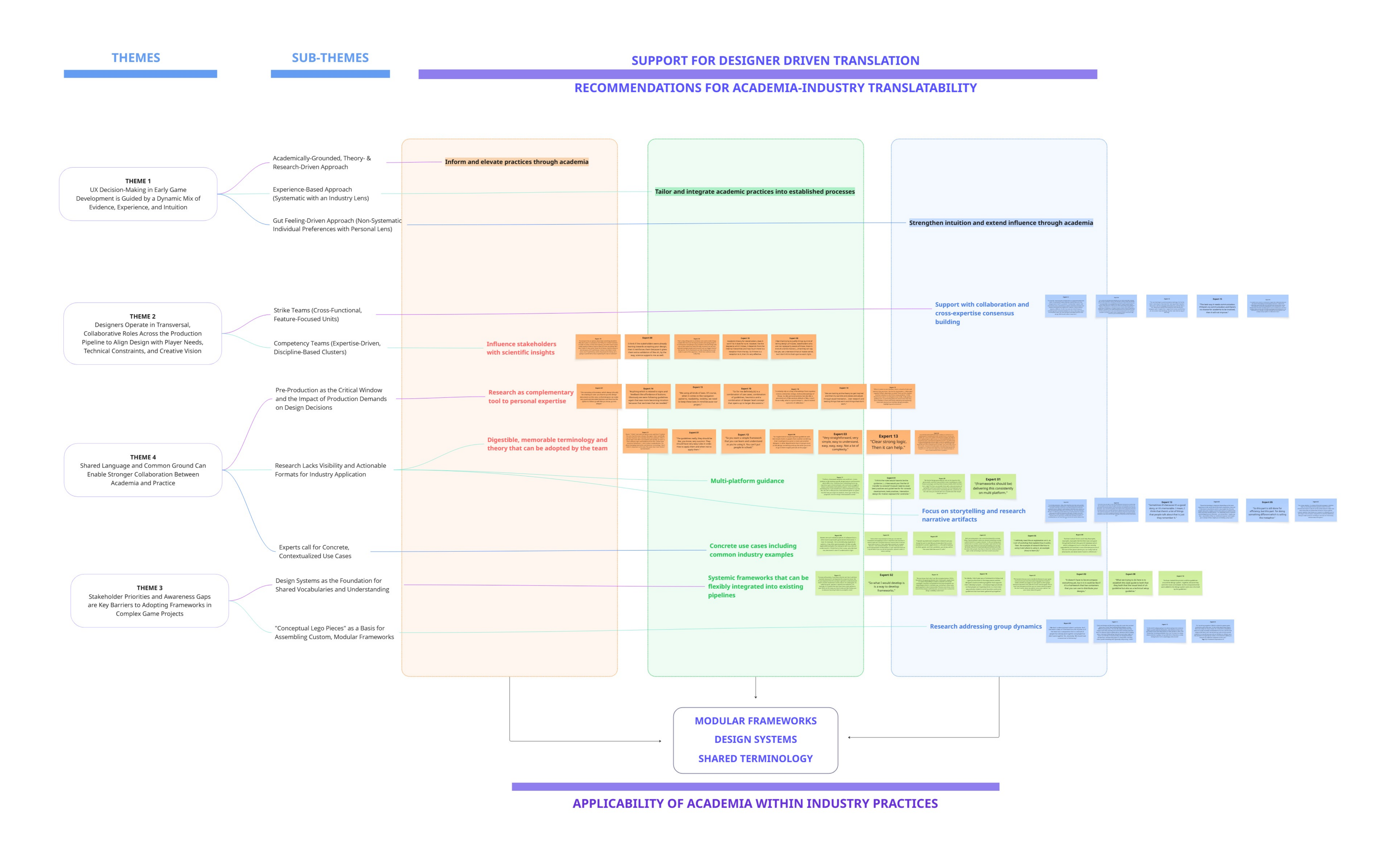}
    \caption{A mapping of our themes, sub-themes supported by participant quotes to derive our model.}
    \label{fig:mappingofthemes}
    \Description{A diagram made up of several flowcharts. Each of the four main themes is expanded into its sub-themes, and each sub-theme is paired with representative participant quotes. The flowcharts show how these themes and quotes connect to, and were used to derive, the Adaptive Design Judgment (ADJ) model.}
\end{figure}

To clarify the provenance of the observations in Figure~\ref{fig:mappingofthemes}, we distinguish findings that are empirically grounded in our data from the higher-order model that represents our interpretive synthesis. The themes and sub-themes were derived from our coded interview data. Participant quotes are denoted to support for designer driven translation and recommendations for academia-industry translatability. Accordingly, the connections in the figure denote our analytic reasoning---how data-driven sub-themes were aggregated into themes and then synthesized into the model---rather than relationships observed directly in the data. This separation makes explicit which claims rest on participant evidence and which reflect our theoretical framing, allowing each to be evaluated on its appropriate basis. A full resolution of all the figures are included in our supplementary material.